\documentclass[11pt]{article}

\usepackage[authoryear]{ natbib }
\usepackage{amsmath}
\usepackage{amssymb}
\usepackage{relsize}
\usepackage{enumerate}
\usepackage{cases}
\usepackage{verbatim}
\usepackage[toc,page]{appendix}
\usepackage[font=small,labelfont=bf]{caption}
\usepackage{colortbl}
\usepackage{ graphicx, psfrag }
\usepackage{multirow}
\usepackage[table,xcdraw]{xcolor}
\usepackage{setspace}
\usepackage{float}

\begin{document}

\title{\mbox{} \vspace*{-1.0in} \\ \textbf{Causal Inference in Repeated Observational Studies: A Case Study of eBay Product Releases}}

\author{
  Vadim von Brzeski \\
  \textit{eBay Inc. and University of California, Santa Cruz} \\
\and
  Matt Taddy \\
  \textit{Booth School of Business, University of Chicago} \\
\and
  David Draper \\
  \textit{University of California, Santa Cruz}
}
\date{}
\maketitle

\mbox{} \vspace*{-0.6in}

\begin{abstract}

Causal inference in observational studies is notoriously difficult, due to the fact that the experimenter is not in charge of the treatment assignment mechanism.  Many potential confounding factors (PCFs) exist in such a scenario, and if one seeks to estimate the causal effect of the treatment on a response, one needs to control for such factors.  Identifying all relevant PCFs may be difficult (or impossible) given a single observational study.  Instead, we argue that if one can observe a sequence of similar treatments over the course of a lengthy time period, one can identify patterns of behavior in the experimental subjects that are correlated with the response of interest and control for those patterns directly.  Specifically, in our case-study we find and control for an \emph{early-adopter effect}: the scenario in which the magnitude of the response is highly correlated with how quickly one adopts a treatment after its release.

We provide a flexible hierarchical Bayesian framework that controls for such early-adopter effects in the analysis of the effects of multiple sequential treatments. The methods are presented and evaluated in the context of a detailed case-study involving product updates (newer versions of the same product) from eBay, Inc.  The users in our study upgrade (or not) to a new version of the product at their own volition and timing.  Our response variable is a measure of \emph{user actions}, and we study the behavior of a large set of users (n = 10.5 million) in a targeted subset of eBay categories over a period of one year. We find that (a) naive causal estimates are hugely misleading and (b) our method, which is relatively insensitive to modeling assumptions and exhibits good out-of-sample predictive validation, yields sensible causal estimates that offer eBay a stable basis for decision-making.

\end{abstract}

\doublespacing

\section{Introduction} \label{introduction}

Causal inference is a complex problem with a long history in statistics.  Its general setup is as follows: given an observable response $Y$, a measurable treatment $Z$, a set of $n$ subjects $i=1,\dots, n$, partitioned into distinct \emph{treatment} ($T$) and \emph{control} ($C$) groups, how much of the observed response was \emph{caused} by the treatment?  In the case of binary treatments, the \emph{potential outcomes} approach \citep{neyman93,rubin74} defines for each subject $i$ two potential outcomes: the response of the subject under treatment $Y_i(Z_i = 1) \equiv Y_i(1)$, and the response of the subject under no-treatment (control) $Y_i(Z_i = 0) \equiv Y_i(0)$.  However, for any individual subject $i$, we cannot observe both outcomes $Y_i(0)$ and $Y_i(1)$, hence the designation \textit{potential} outcomes.  Therefore, the \textit{fundamental problem of causal inference} \citep{holland86} is that to estimate the causal effect of a treatment, we need to compare the two potential outcomes for each individual, namely $[ Y_i(1) - Y_i(0) ]$, but we get to observe only one of those quantities: either $Y_i(1)$ \emph{or} $Y_i(0)$.  

This task is further complicated in \emph{observational studies}. Unlike randomized controlled trials \citep{Fisher1935}, observational studies are characterized by the fact that the experimenter is not in charge of the treatment assignment mechanism.  A treatment event occurs at some point in time, and data are collected on subjects before and after the treatment.  Such a scenario makes it quite likely that many \emph{potential confounding factors} (PCFs) exist.  PCFs are attributes of the subjects (usually covariates) that are correlated with \emph{both} the treatment assignment and the response, and their existence leads to biased estimates of the causal effect unless they are adjusted/controlled for in some way.  Thus, \emph{in addition to} modeling potential outcomes, causal inference in observational studies \emph{requires the discovery of all PCFs} that could have a bearing on valid estimation of the causal effect.  
 
Without loss of generality, let us imagine an observational study in which the response is some measure of user activity (e.g., miles jogged, items bought, ads clicked), and where the availability of a treatment is announced at some point in time.  Users take advantage (or not) of the treatment at their own volition over the subsequent days or weeks, and the response of each user is recorded over time.  Furthermore, suppose that (a) the majority of users who adopt the treatment at all do so in a relatively short time period after its release, and (b) those users who are the earliest adopters exhibit a \emph{higher average response} compared to those who wait longer to try the treatment.  In other words, the \emph{waiting time} to adopt the treatment is (negatively) correlated with the treatment and the response, making it a PCF.  We refer to this situation as the \emph{early-adopter effect}: the overall response is a (confounded) combination of the actual effect of the treatment and the effects of characteristics associated with being an early adopter.

Situations where the early-adopter effect occurs arise with some frequency. For example, suppose that a new diet and exercise plan is offered to the general public for free by a public-health agency. It is a well-known fact in public health that, ironically, those who voluntarily adopt measures to improve their health are precisely the people who need such measures the least, namely people who are already health conscious. Comparing the health status, (say) six months after the plan is offered, of those who chose to use it and those who did not will confound the effect of the plan with the early-adopter effect.

This brings us to the major contribution of this paper. We demonstrate that in observational studies where the early-adopter effect exists, it is difficult to obtain a reasonable estimate of the treatment effect (on the treated) when one only considers a \emph{single treatment event}.  However, we also show that the task is made considerably easier when one studies a \emph{sequence of similar treatments} over an extended period of time.  In the single treatment event scenario, one's only option is to discover (typically static) user attributes that control for the early-adopter effect; in other words, what is it about a user that makes him or her an early adopter?  This may be a difficult or impossible task if little or no data (e.g., demographic information) is available on the users.  On the other hand, given a sequence of similar treatments, the problem is greatly simplified if we assume that the (unknown) early-adopter behavior is relatively consistent from one treatment event to the next.  In such a scenario, we do not need to know the true characteristics (true PCFs) that make a user an early adopter.  Instead, we simply include a set of (indicator) covariates that encode a user's waiting time into our models, and thus account for the early-adopter portion of the total response, leading to a less biased estimate of the treatment effect.  Given a sequence of $K$ treatments $T_1, \dots, T_K$, our approach makes the following two assumptions (which can be verified by exploratory data analysis; see below):

\begin{itemize}

\item  

\textbf{Early Adopter Effect}: The average response per user subsequent to a treatment release should follow a similar \emph{decaying pattern} regardless of the particular treatment.  If this is not the case, the early-adopter effect may not exist at all.  In one extreme scenario, we can imagine the early-adopter effect for some treatments, and a \textit{late-adopter effect} for other treatments in which the average response shows an \emph{increasing} pattern following its release.  In such a scenario, we would need to consider including 
second-order interactions into our models to account for this.

\item 

\textbf{Identifiability}: The users' treatment adoption pattern (i.e., the specific timing with which each user adopts a new treatment after its availability) should differ appreciably between treatments.  This allows us to have an identifiable model.  In the extreme scenario, if each treatment shows the exact same pattern of adoption across all users, we will have collinear columns in our design matrix, resulting in a non-identifiable model. We return to this subject in Section \ref{results}.

\end{itemize}  

The remainder of the paper is organized as follows.  In Section \ref{definitions}, we review the standard estimators of treatment effects found in the observational-studies literature and describe the estimator we will be using in our work.  We also describe the exact nature of the causal inference problem at eBay.  Section \ref{models-and-matrices} details our models, design matrices, and counterfactual computations.  Our results (causal effect estimates) are given in Section \ref{results}.  We describe our model validation approach in Section \ref{model-selection}, where we check our assumptions and investigate the out-of-sample performance of our models.  We conclude in Section 6 with a summary of our major results. \vspace*{-0.2in}

\section{Problem Statement and Definitions} \label{definitions}

\mbox{} \vspace*{-0.7in}

\subsection{Case Study: eBay Product Releases} \label{product-releases}

Our case study deals with a sequence of observational studies at eBay Inc., in which analysts attempted to infer the causal effect of new versions (releases) of a specific software product, henceforth referred to as the \emph{Product}, on aggregate \emph{User Actions} with said Product (the true response and the true product are not disclosed for confidentiality reasons).

The exact nature of the Product is not important; however, it possesses a number of characteristics that are relevant to our study.  First, newer versions (upgrades) of the Product are released on a semi-regular basis, with releases happening on the order of $6-12$ weeks apart on average.  Second, once a new version of the Product is released and becomes available to the general public, users adopt (upgrade to) the new version \emph{at their own volition and timing}.  The new version of the Product is not an \emph{en masse} replacement of the previous version; instead, users choose to upgrade to it or not.  Some users upgrade immediately when (or shortly after) the version becomes available: we will refer to these users as \textit{early adopters}; some users never upgrade and continue to use the same version of the Product throughout our study.  This rolling treatment setting with user Product choice is precisely what makes this an observational study.

\begin{figure*}[t!]

\centering

\includegraphics[scale=0.8]{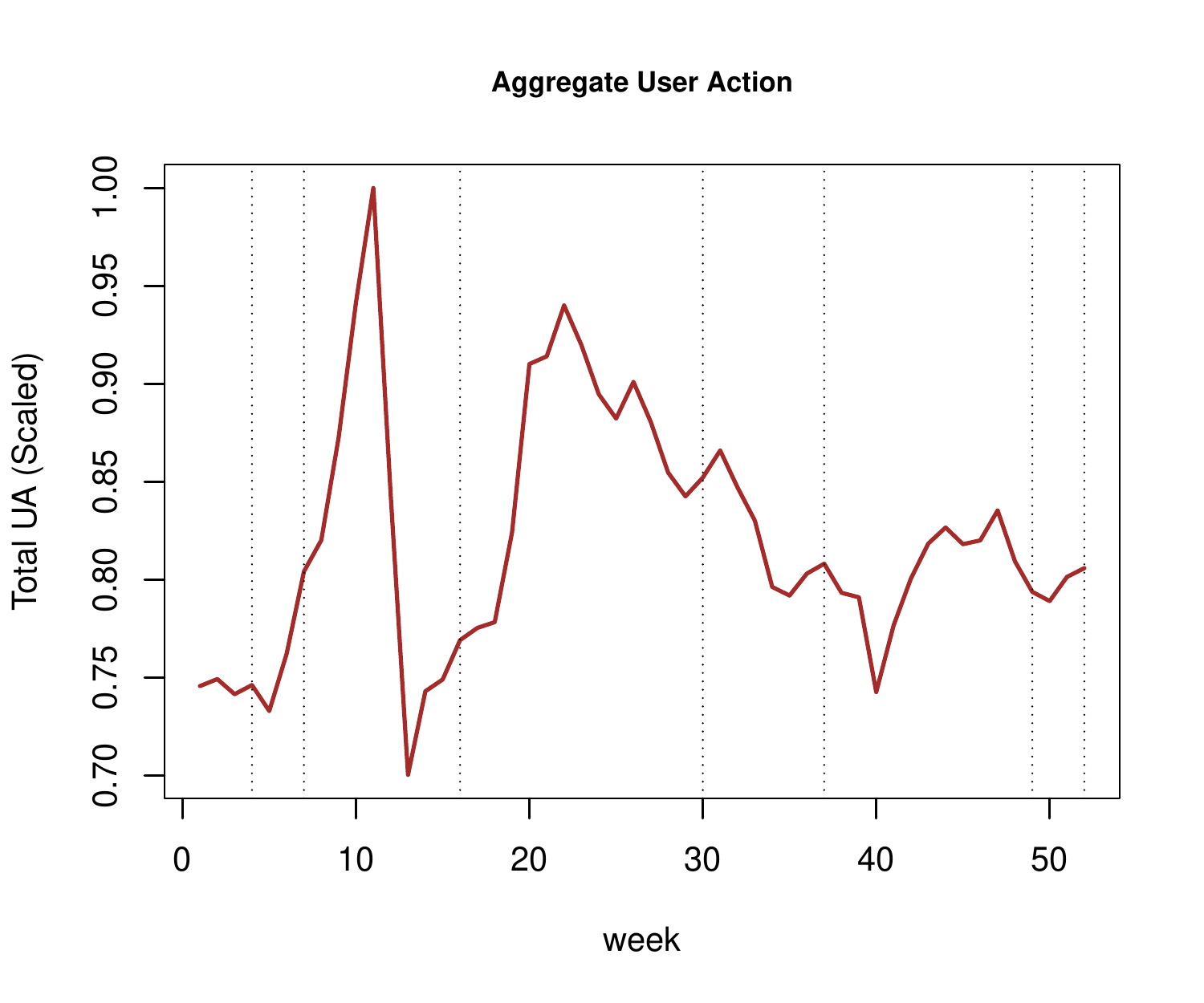}

\vspace*{-4mm}

\caption{\textit{Aggregate (scaled) UA for 10.5M users over the 52 week period we study. The vertical lines indicate the weeks of new Product version releases.}}

\label{totalgmb}

\end{figure*}

For the purposes of this paper, \emph{User Action} is a normalized, non-negative, unit-less quantity reported in \textit{user-action units} (UAs).  Higher aggregate values of UA imply higher (aggregate) levels of satisfaction with the Product by the users in our study.
A graph of weekly aggregate UA over our 52-week study period is shown in Figure \ref{totalgmb}.  The dashed vertical lines in Figure \ref{totalgmb} indicate weeks of Product releases: there were 7 unique releases (treatments) in our 52-week time window.  Each release corresponds to a new version (upgrade) of the Product, e.g., Version 8 to Version 9.  

We would like to estimate what the UA graph in Figure \ref{totalgmb} would look like in the following counterfactual setting.  Suppose that we take two consecutive version releases, say $v_1$ and $v_2$, released on weeks $t_1$ and $t_2$, respectively, and we take $v_2$ as the \emph{counterfactual version} (the one whose causal effect we want to estimate).  Now, suppose that instead of releasing $v_2$ in week $t_2$, \emph{eBay instead releases $v_1$ again but labels it as} ``$v_2$''.  One can envision this counterfactual universe as eBay releasing a \textit{placebo} version, which has a new label but is in fact identical in functionality to its predecessor.  We use this counterfactual construction because we are interested not in what would have happened had a release never occurred at all; instead we ask what effect the features of the new release had on User Action.  
It is worth emphasizing that no one can ever know the true counterfactual given this data-gathering method (observational study); we cannot roll back time and roll it forward again in an alternate universe.  

\subsection{Our Approach and Data} \label{approach-and-data}

We approach the observational study problem from a longitudinal perspective and jointly model the sequence of Product releases.
Our dataset consists of the UA response for $\approx 10.5M$ eBay users over an (undisclosed) 52 week period.  The data is aggregated week by week, i.e., $t = 1,\dots,T$, where $T=52$.  For each user $i = 1, \dots, n =$ 10,491,859, we have Product usage data (session logs) broken out by version; i.e., for each week, we know which version of the Product a user had, and if he (she) upgraded mid-week, we know the relative proportion of each version's usage during that week.  A user was included in our study if he (she) was a registered eBay user as of the first day of our study, \emph{and} had at least one Product session logged in our 52-week window.  Note: our response UA is correlated with Product usage (number of sessions logged), but it is not the same as Product usage.  A frequent user (many logged Product sessions) can still have zero UAs logged.

Our dataset contains 11 distinct Versions: 2, 3, 4, 5, \textbf{6, 7, 8, 9, 10, 11, 12} (so designated for confidentiality reasons); the versions in boldface were released during our 52-week window (the others were legacy versions).  We also had some users on versions prior to Version 2; we lump all these into a \emph{pre\_v\_2} category.  This gives us a total of $R=12$ version indicators.

\begin{figure*}[t!]

\centering

\includegraphics[scale=0.9,trim=0cm 0cm 0cm 0cm,clip=true]{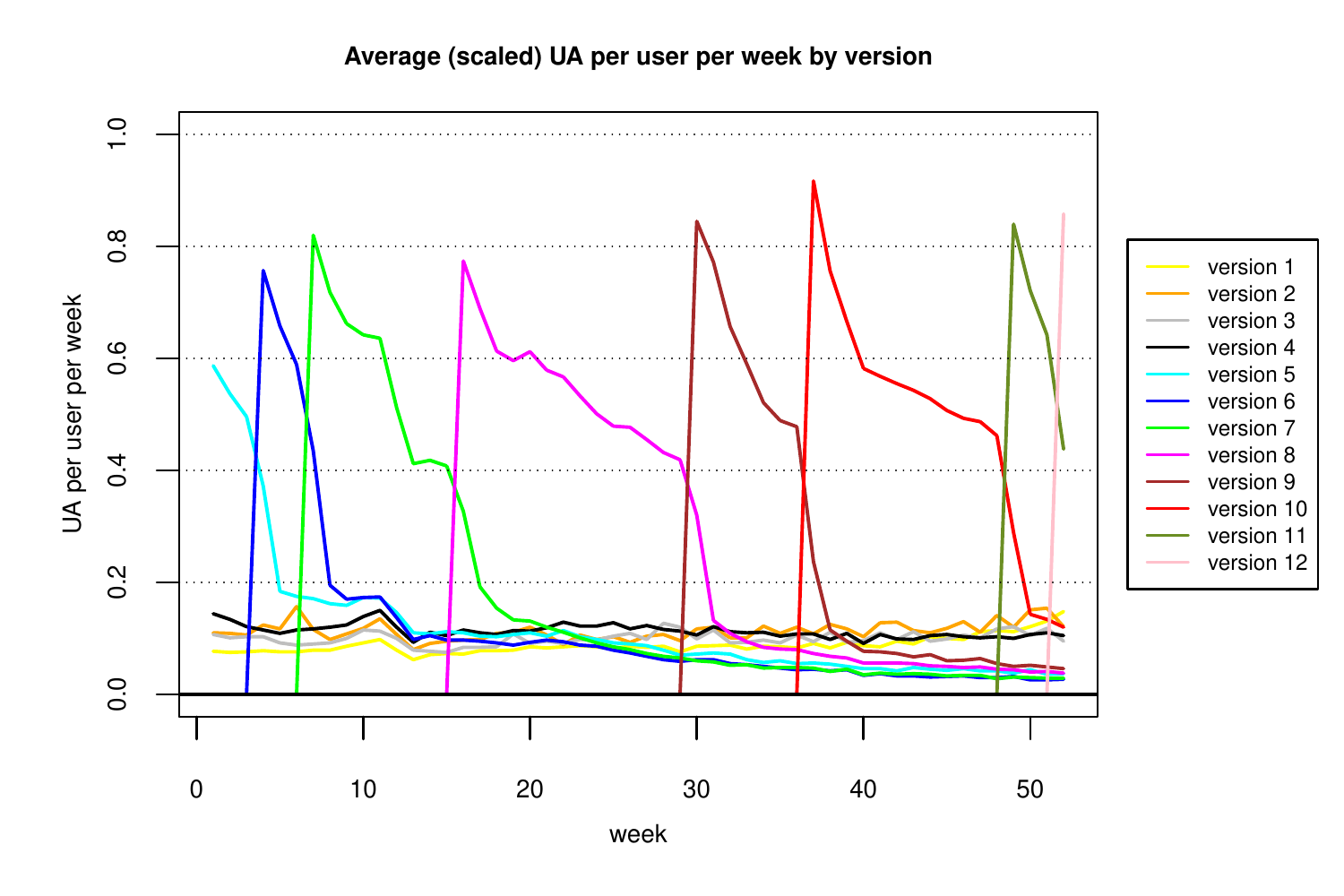}

\vspace*{-4mm}

\caption{\textit{Average (scaled) true UA per user per week for each individual version. This graph shows that the early adopters of a new release have the highest UA average, and that the \underline{early-adopter effect exists} and is quite regular from version to version.}}

\label{avg_gmb_per_user_per_week_true}

\end{figure*}

To determine if our case study exhibits the early-adopter effect, we constructed the graph given in Figure \ref{avg_gmb_per_user_per_week_true}, which shows the average UA per user per week for each individual version.  Two points are made clear by the curves in Figure \ref{avg_gmb_per_user_per_week_true}:

\begin{itemize}

\item

Users who upgrade to a new version in its first weeks of availability are the ones who are the most active on average (measured in UA units): these are the early adopters. Average UA per user declines as more and more late adopters join the ranks and upgrade to the latest version.

\item 

The UA pattern from one version release to another is quite consistent and exhibits a similar decaying pattern for each release, thus confirming one of our earlier assumptions and making it possible to borrow strength across releases. 

\end{itemize}

\subsection{Estimates of Treatment Effects and Assumptions} \label{estimators}

We briefly review the standard estimators for causal effects found in the literature and used in practice \citep{imbens2004nea}, and we discuss the one we chose for our case study. We also discuss some of the assumptions involved when using these estimates of causal effects. 

\begin{itemize}

\item 

\emph{Average Treatment Effect} (ATE): ATE is defined as $E[Y_i(1) - Y_i(0)]$, and it is a measure of treatment effect over the \emph{population}, where the expectation is with respect to the distribution induced by random sampling of observed units from the population. 

\item 

\emph{Sample Average Treatment} Effect (SATE):  SATE is defined as $\frac{1}{n}\sum_{i=1}^n [Y_i(1) - Y_i(0)]$; its computation is similar to that of the ATE, but only for the sample, not the entire population.

\item 

\emph{Conditional Average Treatment Effect} (CATE): CATE is defined as $\frac{1}{n}\sum_{i=1}^n E[Y_i(1) - Y_i(0) | \mathbf{X}_i$ $= \mathbf{x}]$, i.e., the treatment effect on the population conditional on some covariates (PCFs) $\mathbf{X}_i$.

\end{itemize}

The first two treatment estimators (ATE and SATE) do not apply in our case since (a) we are not interested in drawing inferences about a population of eBay users, and (b) we do condition on $\mathbf{X}$ in our models.  However, we also do not target the CATE here because it requires the estimation of two counterfactuals: $Y_{i:Z_i=0}(1)$, the response of the $C$ users if they had been treated, and $Y_{i:Z_i=1}(0)$, the response of the $T$ users had they remained in the control group. Given the data resources in our case study, we are able to reliably place the treated (upgraders) into the control group (non-upgraders), but are not able to reliably predict \emph{who out of the non-upgraders would upgrade} and \emph{when they would upgrade}.  

Therefore, here we employ the Conditional Average Treatment (Effect) on the Treated (CATT) as our measure of causal effect, initially similar to the above estimators but only dealing with the treated group. CATT is defined as:
\begin{align} \label{CATT1}
\text{CATT} &= \frac{1}{n_T}\sum_{i:Z_i=1} E[Y_i(1) - Y_i(0) | \mathbf{X}_i = \mathbf{x}] \, .
\end{align}
Three assumptions are relevant to the quality of CATT as a causal effect estimate.

\begin{itemize}

\item 

\emph{Ignorability assumption} \citep{Rubin1983}: $( Z_i \perp Y_i(0),Y_i(1) | \mathbf{X}_i = \mathbf{ x } )$. This assumption states that if indeed all PCFs $\mathbf{X}_i$ have been controlled for, then treatment assignment $Z_i$ and response $Y_i$ are conditionally independent given the PCFs.  If this is indeed the case, it can be shown that the causal effect estimate will be unbiased \citep{Rubin1983}. This assumption can essentially never be fully verified, because one can never know if one has in fact controlled for \emph{all} confounding factors. 

\item 

\emph{Overlap assumption for CATT} \citep{heckman:ichi:todd:1997}: $\text{Pr}(Z=1 | \mathbf{X}=\mathbf{x}_i) < 1$. This assumption states that when conditioning on some $\mathbf{X}=\mathbf{x}_i$, one cannot have all subjects in the treatment group: there must be some subjects in $C$, else one cannot estimate the effect on the treated using the potential outcomes framework.  This assumption can be verified to some extent, and we do so in Appendix \ref{model-validation-overlap}.

\item 

\emph{Stable Unit Treatment Value Assumption (SUTVA)} \citep{Imbens2015Book}. This assumption states that the potential outcomes for any unit do not vary with the treatments assigned to other units.  In other words, whether a given subject is treated or not has no impact on another subject's response and vice-versa. We assume that SUTVA holds in our scenario because we have been told by eBay Product Managers that the particular eBay product under consideration does not have a viral nature (e.g., an exponential adoption rate).

\end{itemize}

Under the ignorability assumption above,  and for some flexible function (model) $f$,
$E[Y_i(0) | \mathbf{X}_i = \mathbf{x}] = E[Y_i | Z_i = 0, \mathbf{X}_i = \mathbf{x}] = f(0,\mathbf{x})$,
our CATT estimates become
\begin{align} \label{catt2}
\text{CATT (per treated user)} &= \frac{1}{n_T}\sum_{i:Z_i=1} [Y_i - f(0,\mathbf{x}_i)] = \frac{1}{n_T}\sum_{i:Z_i=1} [Y_i - \hat{\mathbf{y}}_i^{CF}] \, , 
\end{align}
and we define the \emph{CATT Causal Ratio (CCR)} as
\begin{align} \label{catt2lift}
\text{CATT Causal Ratio (CCR)} &= \dfrac{\sum_{i:Z_i=1} f(0,\mathbf{x}_i)}{\sum_{i:Z_i=1} Y_i}= \dfrac{\sum_{i:Z_i=1} \hat{\mathbf{y}}_i^{CF}}{\sum_{i:Z_i=1} Y_i} \, .
\end{align}
Simply put, the CCR is the ratio of the aggregate $\hat{\mathbf{y}}_i^{CF}$ to the aggregate $\mathbf{y}_i$ for the treated group.  All results below are reported in terms of CCR.  Note that if 
$\sum_{i:Z_i=1} \hat{\mathbf{y}}_i^{CF} = \sum_{i:Z_i=1} Y_i$, then CCR=$1$, which means that the treatment had no causal effect on the response of the treated; CCR values less than 1 suggest that the effect caused by the Product release was to increase UA (user satisfaction) on average.

Extensive discussions with relevant eBay experts identified a strong source of information about CCR external to our data: having launched a number of releases of the Product in the past without dramatic apparent positive or negative effects on important indicators correlated with UA, company experts were highly skeptical of CCR values far from 1. We use this (prior) information informally as a kind of baseline in what follows.

\subsection{Related Work} \label{related-work}

Our approach is fundamentally model-based, but other methods for causal inference in observational studies of course exist.
As mentioned above, one leading approach to estimating causal effects
is via a comparison of potential outcomes. However, the problem is complicated by the presence of unknown PCFs, and thus the problem boils down to controlling for such PCFs when comparing responses in $T$ and $C$.  

One of the most widely used techniques relies on \emph{matching} \citep{Rubin1973} treated and control subjects on the hypothesized PCFs (covariates), with the goal of achieving a \emph{balance in the covariate distributions} in the $T$ and $C$ groups \citep{rosenbaum1985mj}.  Once a set of covariates is identified, matching algorithms attempt to find the closest match to a treatment subject in the control group, in a particular sense of closeness. Having identified the best control subject for each treatment subject, the algorithm computes the average difference between the pairs of treatment and control subjects. For a good review of matching techniques, see \citet{stuart2010}.

One way to define the concept of closeness is through \emph{propensity score} matching \citep{Rubin1983}. The propensity score for a subject $i$ is defined as the probability of receiving the treatment given the observed covariates. There are two important properties of propensity scores. First, at each value of the propensity score, the distribution of the covariates defining the score is the same in the $T$ and $C$ groups, i.e., they act as balancing scores.  Second, if treatment assignment is ignorable given the covariates, it is also ignorable given the propensity score.  Thus to compute the causal effect, one can compare the mean responses of treated and control subjects having the same propensity score. However, the above two properties only hold if one has found the \emph{true} propensity-score model: a poor estimate of the \emph{true} propensity score will again lead to biased causal effect estimates \citep{kang2007ddr}. Besides matching on the propensity score, other techniques involve using the propensity score in subclassification \citep{rose:rubi:redu:1984}, weighting \citep{rosenbaum:1987e}, regression \citep{heckman:ichi:todd:1997}, and/or combinations of the above \citep{rubi:thom:comb:2000}. Bayesian analyses using propensity scores also exist \citep{McCandless2009}.

\emph{Instrumental variables} (IV) \citep{angrist1996} also have a long history, and are widely used in econometrics as a way to approach unbiased causal estimates in the presence of PCFs.  The key idea behind IV is that if we can find an \textit{instrumental variable} $z$ with the property that it affects the response $y$ \emph{only through its effect} on a PCF $x$ and is uncorrelated with the error, then we can still estimate the effect in an unbiased fashion.  The issue is that such variables, whose only impact on the response is \emph{indirectly} through another covariate, are not easy to find in most situations.

The above approaches do not rely on any specific model of the data; they compare mean responses between specially constructed samples of subjects from $T$ and $C$.  Model-based approaches (such as ours in this paper) attempt to jointly model the treatment and the response in a flexible way so that the unknown counterfactual potential outcomes can be estimated (predicted) by the model. The models are typically linear regression models of the response, but can also be sophisticated non-parametric models (e.g., decision trees) \citep{Hill07bayesiannonparametric, Karabatsos2012925}.  A recent method utilizes a Bayesian time-series approach and a diffusion-regression state-space model to estimate the causal effect of an advertising campaign \citep{brodersen2015}; this approach is closest in spirit to our methodology, but it analyzes the effect of only a single intervention.

\subsection{Previous eBay Causal Estimates} \label{previous-estimates}

We now describe a previous approach at eBay to the above causal inference problem; the method focused on analyzing the causal effect of one release at a time.  A pool of users was selected based on activity logs in a $\pm \, 2$ week window around the release in question (the counterfactual release).  The pool of users was then divided into $T$ and $C$ groups based on their version usage during the pre-release and post-release window. The UA for both groups was computed in a 2 week window before the release, and in a 2 week window that started 3 days (burn-in) after the day of Product release.  The results are given in Table \ref{previous-results1}. A simple (\textit{unadjusted}) estimate using the means of the treatment and control groups shows CCR values in excess of $1.30$ (enormous in relative terms), naively implying a huge effect of the release on customer satisfaction. The \textit{PCF-adjusted} CCR (1.10 to 1.14) was computed using the CATE version of the CCR estimator described above, using a regression model that included hypothesized PCFs as covariates.  No one in the organization believed the unadjusted estimates, and although the 
PCF-adjusted numbers were more reasonable, no one believed them either because they were still large in relative terms.  Therefore, a new approach was necessary.  (Note that the results were similar for \emph{two} releases, suggesting that each release had a significant impact on UA, further eroding the credibility of this approach.)

\begin{table}[t!]

\centering

\begin{tabular}{c|c|c|c|c}

Version & $n$ Treated & $n$ Control & Unadjusted CCR & PCF-Adjusted CCR \\ 

\hline \hline

5 & 3638K & 561K & 1.35 & 1.14 \\ 

\hline 

7 & 3838K & 704K & 1.32 & 1.10 \\ 

\end{tabular}

\vspace*{0.1in}

\caption{\textit{Previous (simple) attempts at answering the causal effect question led to unrealistic causal effect estimates. \underline{Unadjusted} refers to a simple comparison of means (null model); \underline{PCF-adjusted} refers to regression models with a variety of PCFs as covariates, and is based on the CATE estimator.}}

\label{previous-results1}

\end{table}

\section{Model and Design Matrices} \label{models-and-matrices}

We fit our data using variations of a Bayesian hierarchical (mixed effects) model with a Gaussian error distribution (see Section \ref{sensitivity} for  sensitivity analyses of this choice of model class).
Many of the models we examined include auto-regressive (AR) terms of different orders $p$. In such cases, we use the standard conditional-likelihood approach \citep{reinsel13} to building the likelihood function with AR terms; this is justified in our case because (a) our outcome variable, with a reasonable number of AR lags, is essentially stationary, and (b) the modeling has the property that the $X$ matrix, when the AR model is estimated via regression, is invertible.  This permits us to regard the $\mathbf{f}_i$ matrix for each user $i$ as a matrix of fixed known constants in the models below.

The dimensions of all the quantities listed below are as follows, where $p$ denotes the AR order.

\begin{itemize}

\item 

$\mathbf{y}_i$: a $(T-p)$ by $1$ vector of user $i$'s response (UA);

\item 

$\boldsymbol{\beta}_i$: a $d$ by $1$ vector; in random effects models, $d$ is the length of the random effects coefficients vector, and includes the AR coefficients;

\item 

$\mathbf{f}_i$: a $(T-p)$ by $d$ matrix of constants and lagged $y_i$ values (see Section \ref{f_i_matrix});

\item 

$\mathbf{W}_i$: a $(T-p)$ by $(T-p)$ matrix of fixed known constants (typically week indicators); and

\item 

$\boldsymbol{\gamma}$ : a $(T-p)$ by $1$ vector of coefficients of the fixed effects.

\end{itemize}

\label{m004def}

Our primary working model is a mixed-effects hierarchical model with Gaussian error. For user $i = 1,\dots,n = $10,491,859, the model is as follows:
\begin{align} \label{primary-model}
\mathbf{y}_i & = \mathbf{f}_i \, \boldsymbol{\beta}_i + \mathbf{W}_i \, \boldsymbol{\gamma} + \boldsymbol{\varepsilon}_i \nonumber \\
( \boldsymbol{\beta}_i \, | \, \boldsymbol{\mu},\mathbf{\Sigma} ) & \sim \textbf{N}(\boldsymbol{\mu},\mathbf{\Sigma}) \nonumber \\
( \boldsymbol{\varepsilon}_i \, | \, \nu ) &\sim \textbf{N}(\mathbf{0},\nu\textbf{I}_{T-p}) \nonumber \\
\boldsymbol{\mu} &\sim \textbf{N}(\mathbf{0}, \kappa_{\mu} \textbf{I}_d) \nonumber \\
\boldsymbol{\gamma} &\sim \textbf{N}(\mathbf{0}, \kappa_{\gamma} \textbf{I}_{T-p}) \nonumber \\
\nu &\sim \text{Inv-Gamma} \left( \frac{ \epsilon }{ 2 }, \frac{ \epsilon }{ 2 } \right) \nonumber \\
\mathbf{\Sigma} &\sim \text{Inv-Wishart}_{d+1}(\textbf{I}) 
\end{align}

This model assumes that each Product version affects all users \emph{differently}, i.e., the model treats all users in a \emph{heterogeneous} fashion, and allows room for homogeneous fixed-effects common to all users in the $\textbf{W}_i$ matrix.
We assume the error distribution to be Gaussian.  This is quite possibly incorrect for individual users, but can still lead to a model that performs well in the aggregate; section \ref{sensitivity} shows that our results are insensitive to an alternative non-parametric error specification.

We employ diffuse (yet proper) priors for $\boldsymbol{\mu}, \boldsymbol{\gamma}$, and $\nu$, namely $\kappa_{\mu} = \kappa_{\gamma} = 10^6$, and $\epsilon = 0.001$.
For the prior on the unknown covariance matrix $\mathbf{\Sigma}$, we choose a diffuse proper prior distribution \citep{bda3} which has the nice feature that each single correlation in the $\mathbf{\Sigma}$ matrix has marginally a uniform prior distribution.  We fit the above mixed-effects model using MCMC, and all full conditional distributions are available in closed form (see Appendix \ref{mcmc-hier-eqn}). Sensitivity analyses not presented here demonstrated that reasonable variations in the hyper-parameters of the diffuse priors had negligible effects on the results, which is to be expected with $n$ in excess of 10 million.

\subsection{$\textbf{f}_i$ Matrix} \label{f_i_matrix}

For each user $i$, the design matrix $\textbf{f}_i$ contains three sets of covariates: (a) the version (treatment) indicators, (b) the PCFs that encode waiting time to adopt the latest version, and (c) other user covariates.  We detail each of these below. 

\subsubsection{Version (Treatment) Indicators}
The $R=12$ version indicator columns of $\textbf{f}_i$ denote which specific version (treatment) user $i$ had installed during each of the $T=52$ weeks.  In detail, $\mathbf{f}_i^{version} = [\mathbf{x}_{i,1}^{'},\mathbf{x}_{i,2}^{'}, \dots, \mathbf{x}_{i,R}^{'}]^{'}$, 
where $r = 1,\dots,R = 12$ is the number of unique Product versions in the study.  Each indicator column $\mathbf{x}_{i,r}$ encodes the weeks user $i$ had version $r$.  In the vast majority of cases, $\mathbf{x}_{i,r}$ only contains $0/1$ values; however, we allow for fractional entries in cases where a user upgraded to a new version midweek. We compute such fractional usage using session data for a given week.

\subsubsection{Waiting Time PCFs} \label{waiting-time-pcfs}

To control for the early-adopter effect mentioned above, we construct 14 binary indicator variables called \textit{$n$-weeks-past-release} indicators.  For a given user $i$ in a given week $t$, we calculate how long ago the current latest version was released, relative to the given week $t$.  For instance, suppose that the current latest version was shipped in week $t_1$, and the given week is $t$; then $n$-weeks-past-release($t$) $ = ( t - t_1 )$.  We then set the $(t-t_1)$-th indicator variable to 1. There are 14 such indicators because that is the maximum number of weeks between consecutive releases. Table \ref{nweeks-sample} presents an example of this calculation for a single user.

\begin{table}[t!]
\setcounter{MaxMatrixCols}{20}
\begin{align}
\begin{bmatrix}
t&0\_wks&1\_wks&2\_wks&3\_wks&4\_wks&5\_wks&6\_wks&7\_wks&8\_wks&...\\
\hline
0& . & . & . & . & . & . & . & . & . & ... \\
1& . & . & . & . & . & . & . & . & . & ... \\
2& . & . & . & . & . & . & . & . & . &  ... \\
3& . & . & . & \textbf{1} & . & . & . & . & . & ... \\
4& . & . & . & . & \textbf{1} & . & . & . & . & ... \\
5& . & . & . & . & . & \textbf{1} & . & . & . & ... \\
6& . & . & . & . & . & . & \textbf{1} & . & . & ... \\ \hline
7& . & . & . & . & . & . & . & . & . & ... \\
8& . & \textbf{1} & . & . & . & . & . & . & . & ... \\
9& . & . & \textbf{1} & . & . & . & . & . & . & ... \\
10& . & . & . & \textbf{1} & . & . & . & . & . & ... \\
\vdots & & & & & .... & & & & & \vdots \\
\end{bmatrix} 
\end{align}
\caption{\textit{Example of a user's \underline{$n$-weeks-past-release} indicator columns $\textbf{f}_i^{waiting\_time}$.  The ``.'' entries represent $0$.  The horizontal lines indicate weeks of new version releases.  This user waited 3 weeks to upgrade to the version released in week $0$ ($t=0$), and waited 1 week to upgrade to the version released in week $7$ ($t=7$).}}
\label{nweeks-sample}  
\end{table}

\subsubsection{Other Covariates}

Looking at all of our 10.5M users, we have approximately 2.83M users whose first recorded Product usage was during our 52 week window.  (This is not to say these users had \emph{never} used the Product before, but we did not find a record of them using the Product in the 12 months prior to the start of our study).  Thus we include a binary indicator covariate called \texttt{virgin\_user} to denote those users who appeared to use the Product for the first time ever in our study window. We also add a covariate that captures a user's long term behavior, namely the six-month rolling average of UA over \emph{all of eBay's products}, not just using the Product in the study. Finally, in order to control for a user's behavior during the one week he or she upgrades, we include a binary indicator covariate (\texttt{upgrade-week}) for the particular week in which an upgrade occurs.

\subsection{$\textbf{W}_i$ Matrix} \label{w-matrix}

Our initial exploratory (\textit{flat}) model assumed that each Product version affects all users equally, i.e., the model treats all users in a \emph{homogeneous} fashion by having a single $\boldsymbol{\beta}$ parameter instead of the $\boldsymbol{\beta}_i$ random effects in model (\ref{primary-model}). We initially captured this idea with the following ordinary least squares (OLS) Gaussian model: for user $i = 1,\dots,n = $10,491,859, 

\mbox{}

\vspace*{-0.8in}

\begin{align} \label{flat-model}
\mathbf{y}_i &= \mathbf{f}_i\boldsymbol{\beta} + \textbf{W}_i \boldsymbol{\gamma} + \boldsymbol{\varepsilon}_i \nonumber \\
( \boldsymbol{\varepsilon}_i \, | \, \nu ) &\sim \textbf{N}(\mathbf{0},\nu\textbf{I}_{T-p}) \nonumber \\
(\boldsymbol{\beta}, \boldsymbol{\gamma}, \nu) &\propto 1 \, .
\end{align}
We immediately found it necessary to account for time in some manner.  Models that did not account for time at all, and models that involved a simple linear time variable ($t$), did poorly in fitting the aggregate response.  We discovered that our best models were those that included an effect for each individual specific week of our 52-week period. Therefore, we included $( T - p )$ (where $p$ is the AR order) indicator columns as fixed effects in the matrix $\mathbf{W}_i$ in model (\ref{flat-model}); each $\mathbf{W}_i$ is effectively the identity matrix of dimension $( T - p )$. These indicator variables can be regarded as proxying for changes over time that are exogenous to our study, both internal and external to eBay.

\subsection{Counterfactual $\textbf{f}_i$ Matrix} \label{cf-matrix}

Throughout our work we estimated the counterfactual response $\hat{\mathbf{Y}}_{CF}$, given a certain Product version, which we call the \emph{counterfactual version} (CV). 
As noted in Section 2.1, we estimate the response (UA) if that particular version had not been released, but a placebo version had been released in its place.

When estimating the counterfactual response, we need a counterfactual counterpart to the version indicator columns described above, namely $\textbf{f}_i^{CF\_version}$.  We construct $\mathbf{f}^{CF\_version}_i$ by simply moving the user from the CV to his or her previous version (note that this is user-dependent).  In the $\mathbf{f}^{version}_i$ matrix, this amounts to adding the CV indicator column to the column corresponding to the user's prior version, and then zeroing out the CV column.
The counterfactual for a virgin user is constructed by moving the user to the most recent previous version.

There is a slight twist to computing counterfactual estimates in models that include auto-regressive AR($p$) terms, as many of our models do.  Suppose that we include an AR(1) term in as a random effect.  In this case, we have to make an adjustment in the counterfactual computation during the time period in which the CV was active: the true lagged $y$ values are replaced by their (sequentially) \emph{estimated} lagged $\hat{y}$ values, but only during the period of time during which the CV was employed by the given user.

\subsection{Counterfactual Computation} \label{cf-computation}

The estimate of the counterfactual $\hat{\mathbf{Y}}_{CF}$ response in our hierarchical mixed-effects model is computed as follows.  Given that we have fit the model and run $M$ samples after burn-in, we have the following sets of samples from the posterior distributions above:

\begin{itemize}

\item 

$M$ samples each of $\boldsymbol{\mu}$, $\mathbf{\Sigma}$, $\boldsymbol{\gamma}$, and $\nu$; and

\item 

$\bar{\boldsymbol{\beta}}_i$ : Since we have $n =$ 10.5 million users in our dataset, and finite memory and disk space, we do not store $M$ samples of each user's $d$ dimensional vector $\boldsymbol{\beta}_i$. Instead, we simply store the mean $\bar{\boldsymbol{\beta}}_i$ for each user $i$, where the mean is taken over the $M$ posterior samples.

\end{itemize}

Given the above, we calculate the counterfactual estimates as follows.  
If we are just interested in point estimates, we simply use the point estimates $\bar{\boldsymbol{\beta}}_i$ from the posterior for each user $i$:
\begin{align} \label{cf-estimate-1}
\hat{\mathbf{y}}_{i}^{CF} = \textbf{f}_i^{CF} \bar{\boldsymbol{\beta}}_i + \mathbf{W}_i \bar{\boldsymbol{\gamma}} \, .
\end{align}
To create uncertainty bands around our estimate, we simulate the following.  For each user $i$, we draw $\boldsymbol{\beta}_i^{*}$ from the its full conditional given the true $\mathbf{f}_i$ matrix and the posterior means of the other parameters, and then draw $\hat{\mathbf{y}}_{i}^{CF}$ using the counterfactual $\textbf{f}_i^{CF}$ matrix:
\begin{align} \label{simycf}
\boldsymbol{\beta}_i^{*} \sim p(\boldsymbol{\beta}_i | \mathbf{y}_i, \mathbf{f}_i, \bar{\boldsymbol{\mu}}, \bar{\mathbf{\Sigma}}, \bar{\nu}, \bar{\boldsymbol{\gamma}}) \nonumber \\
\hat{\mathbf{y}}_{i}^{CF} \sim \textbf{N}(\textbf{f}_i^{CF} \boldsymbol{\beta}_i^{*} + \mathbf{W}_i \bar{\boldsymbol{\gamma}}, \bar{\nu}) \, .
\end{align}
We then sum up each user's CF estimate to obtain the aggregate estimate $\hat{\mathbf{Y}}_{CF} = \sum_{i=1}^{n} \hat{\mathbf{y}}_{i}^{CF}$.
Note that our CATT estimates only consider the counterfactual response during the weeks of a release's lifetime, i.e., when it was the latest release on the market.  In the case of Version 9, this period was from week 30 up to and including week 36, and we make no claims about the counterfactual story thereafter, at which point Version 10 comes on the market.  
The reason for this is as follows.  In our counterfactual constructed universe, during the 7 weeks in which Version 9 was the latest version, users were shifted onto the release they had immediately prior to Version 9 (this varied among users, but the majority were on Version 8). When Version 10 was released, users who upgraded to Version 10 in the true universe were upgraded in the CF universe as well, but users who remained on Version 9 in the true universe were retained on Version 8.  In the window where Version 9 was the latest release on the market (the only game in town, so to speak), this is the only choice available to us.  However, when Version 10 replaces Version 9 as the latest release, we cannot be sure those same users who stuck with Version 9 until the end would have also stuck with Version 8 until the end. 

\section{Estimates of Treatment Effects} \label{results}

In order to motivate our methodology and results, we first demonstrate what happens when one models an individual version release in isolation and also ignores the early-adopter effect.  Next we show that more reasonable estimates are achieved when one takes the early-adopter effect into account, and in order to do so, one must model the entire sequence of version releases.  In the following causal effect estimates, we \emph{take Version 9 as our counterfactual version} released in week 30 (and replaced in week 37) in all models initially; once we have settled on the best model, we apply the same CF estimation technique to Version 10 (released in week 37). 

\subsection{Modeling a Single Version in Isolation} \label{isolation}

\begin{figure*}[t!]

\centering

\includegraphics[scale=0.7,trim=0cm 0cm 0cm 0cm,clip=true]{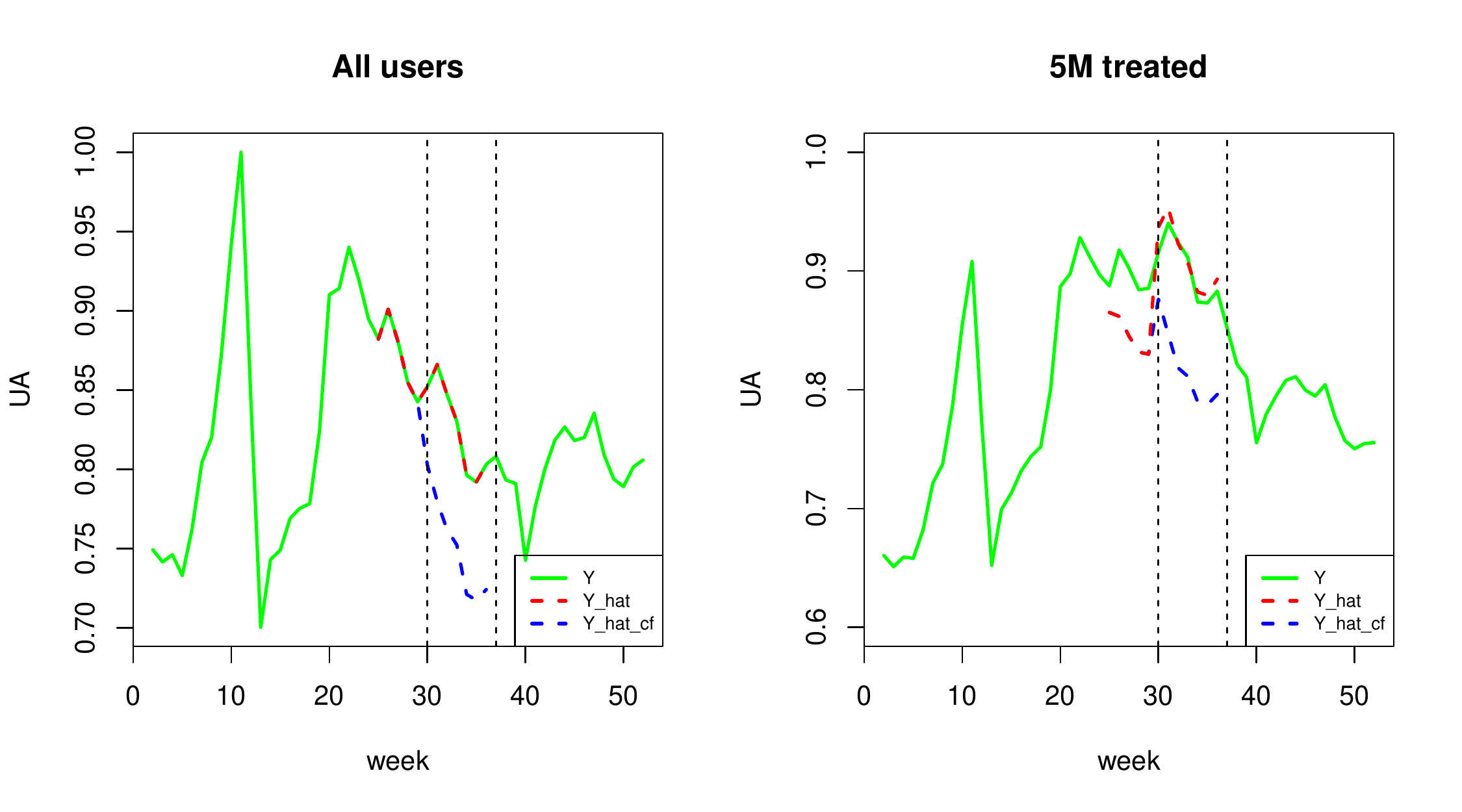}

\caption{\textit{Model $M_R^a$: modeling Version 9 in isolation and ignoring the early-adopter effect (i.e., $n$-weeks-past-release indicators not included).  The estimate of the mean CCR for \textbf{Version 9} is \textbf{0.906}.}}

\label{m004.F-prev.6mo_roll.weeks24-36.ar1.lam0.cf330}

\end{figure*}

As a first step, we ignore the early-adopter effect completely and estimate the causal effect of Version 9 in isolation.  Our first \emph{reduced model} (model $M_R^a$) includes weeks 24 through 36 only, only Versions 1 through 9, and does not include the 14 \textit{$n$-weeks-past-release} indicators.  The results for this are given in Figure \ref{m004.F-prev.6mo_roll.weeks24-36.ar1.lam0.cf330}. The estimate of the mean CCR for Version 9 from model $M_R^a$ is 0.906, which implies that without Version 9, the UA of the treated would have been around 10\% lower in aggregate over weeks 30--36. This is not credible given the informal prior information from eBay Product Managers described in Section 2.3; moreover, the right-hand panel in Figure 3 shows that this model fits the responses of the users in the treatment group poorly in the run-up to the release of Version 9.

\begin{figure*}[t!]

\centering

\includegraphics[scale=0.7,trim=0cm 0cm 0cm 0cm,clip=true]{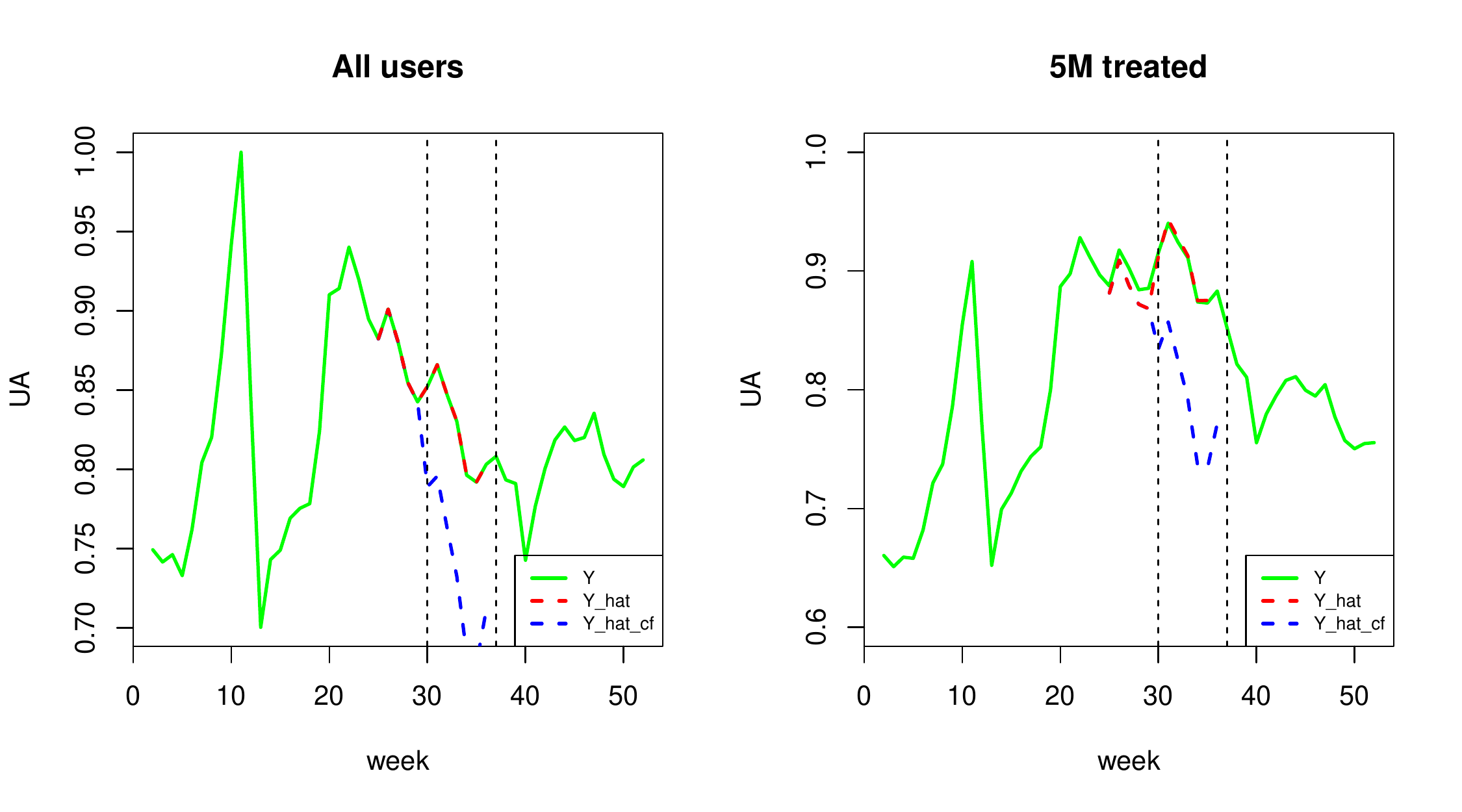}

\caption{\textit{Model $M_R^b$: modeling Version 9 in isolation but naively including $n$-weeks-past-release indicators.  The estimate of the mean CCR for \textbf{version 9} is \textbf{0.878}.}}

\label{m004.F-upweek-and-all14nweeks-random.upgrade-indic.6mo-roll.weeks24-36.ar1.lam0.cf330}

\end{figure*}

Given the unsatisfactory CATT estimate documented in Figure 3, we subsequently modeled the early-adopter effect and included the 14 \textit{$n$-weeks-past-release} indicators in the model.  However, due to computational limitations, we continued with $M_R^a$ and modeled the version in isolation (weeks 24--36). The results for our second reduced model (model $M_R^b$) are given in Figure \ref{m004.F-upweek-and-all14nweeks-random.upgrade-indic.6mo-roll.weeks24-36.ar1.lam0.cf330}.  This did not help: the estimate of the mean CCR for Version 9, 0.878, became even more unrealistic.  

\begin{figure*}[t!]

\centering

\includegraphics[scale=0.7,trim=0cm 0cm 0cm 0cm,clip=true]{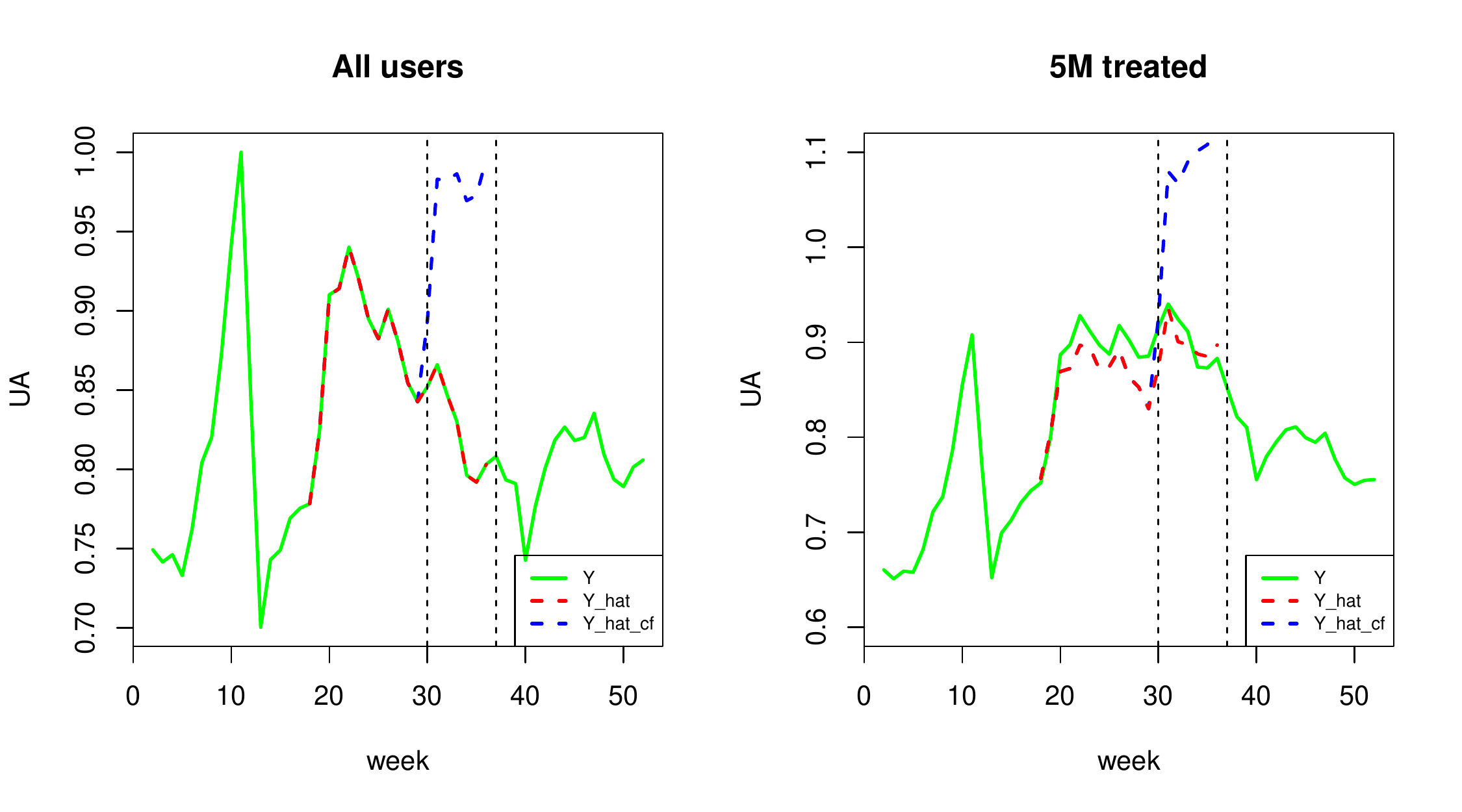}

\caption{\textit{Model $M_R^c$: modeling Version 9 in isolation, naively including $n$-weeks-past-release indicators, and extending the modeling time window to weeks 17--36 (from weeks 24--36 as in Figure \ref{m004.F-upweek-and-all14nweeks-random.upgrade-indic.6mo-roll.weeks24-36.ar1.lam0.cf330}).  The estimate of the mean CCR for \textbf{Version 9} is now \textbf{1.188}, sharply different from the estimates obtained from model $M_R^b$ in Figure \ref{m004.F-upweek-and-all14nweeks-random.upgrade-indic.6mo-roll.weeks24-36.ar1.lam0.cf330}.}}

\label{m004.F-upweek-and-all14nweeks-random.upgrade-indic.6mo-roll.weeks17-36.ar1.lam0.cf330}

\end{figure*}

Still unhappy with the CATT estimates from model $M_R^b$, we concluded that in order to nail down the treatment effect, it was necessary to extend the time window of our analysis.  The earliest week our time window can start is limited by the release week of Version 8, namely week 16, since by definition we are treating Version 9 in isolation.  Thus, our time window for our third reduced model (model $M_R^c$) was weeks 17 through 36, with all covariates unchanged from model $M_R^b$. The results for model $M_R^c$ are given in Figure \ref{m004.F-upweek-and-all14nweeks-random.upgrade-indic.6mo-roll.weeks17-36.ar1.lam0.cf330}.  Surprisingly, model $M_R^c$ shows a much different CCR estimate of 1.188, \emph{larger in magnitude and in a different causal direction}.  Further perturbation of the time window (not shown), namely weeks 21 through 36, yields a similarly large (and non-credible) CCR estimate of 1.35.

It is clear that modeling Version 9 in isolation \emph{and} naively incorporating the early-adopter effect (via the $n$-weeks-past-release indicators) leads to highly volatile estimates of the treatment effect.  CCR estimates in these models are extremely sensitive to the choice of the time window of the model. The reason for this has to do with the \emph{identifiability} of the model.  A reduced model, which considers a release in isolation \emph{and} which includes the $n$-weeks-past-release covariates to account for the 
early-adopter effect, runs the risk of being non-identifiable because \emph{by construction} the treatment indicator column approaches a linear combination of the $n$-weeks-past-release indicator columns: in model $M_R^b$ (weeks 24--36), the correlation between the Version 9 indicator column and the \emph{sum} of the first 7 ``$n$-weeks-past-release'' columns was 0.984. The correlation was computed over all 10.5M users for the 12-week time window (13 weeks minus 1 for AR(1)); the constructed (126M $\times$ 1)-dimensional vectors differed in fewer than 0.7\% of the locations.  Clearly models with such collinear covariates are on the edge of identifiability, leading to wildly varying estimates of parameters and consequently of treatment effects.  Furthermore, this makes it is almost impossible to isolate the early-adopter effect from the version treatment effect.  Nevertheless, the early-adopter effect is real (recall Figure \ref{avg_gmb_per_user_per_week_true}) and must be accounted for.

\subsection{Modeling All Versions Jointly} \label{jointly}

\begin{figure*}[t!]

\centering

\includegraphics[scale=0.7,trim=0cm 0cm 0cm 0cm,clip=true]{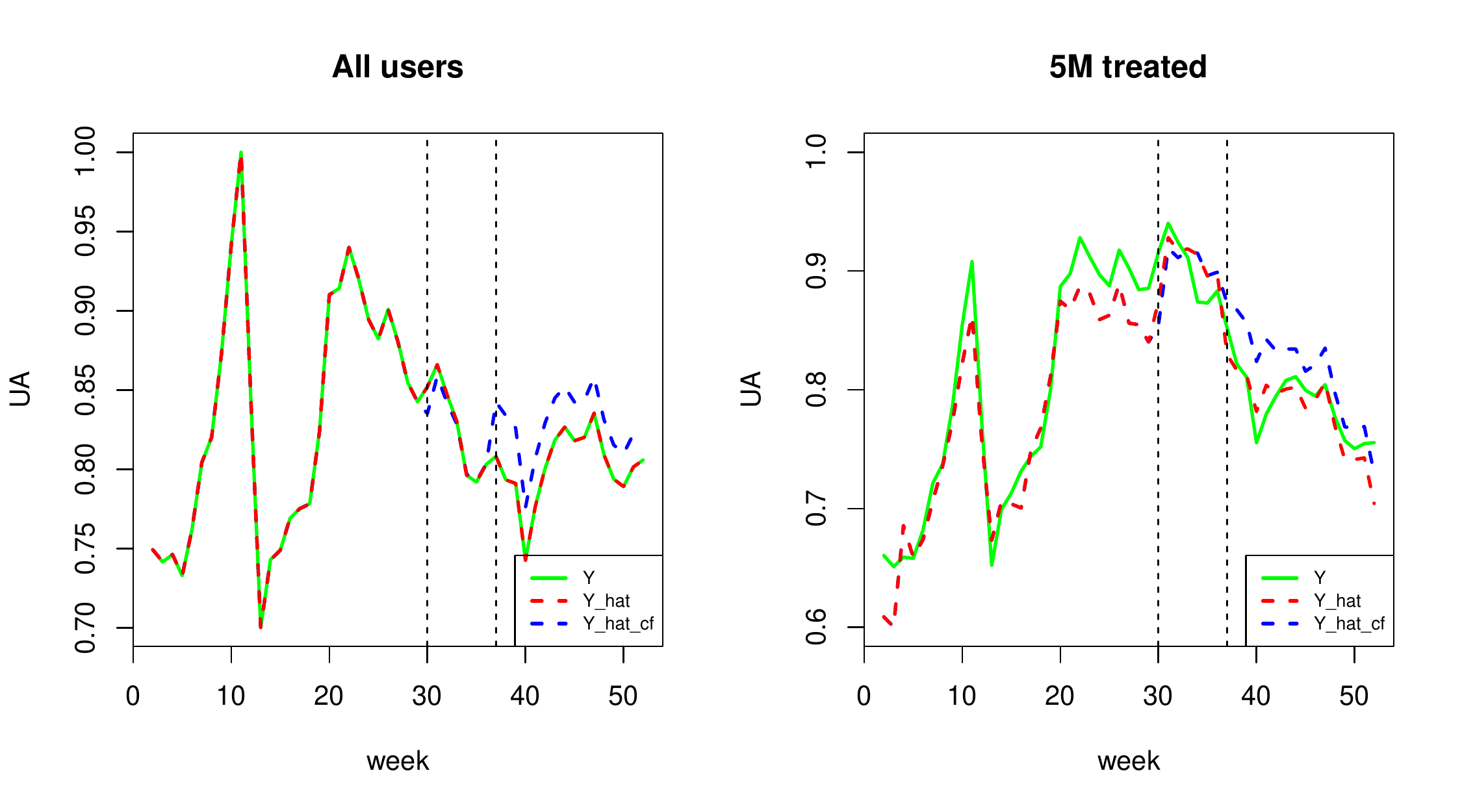}

\caption{\textit{Modeling all releases (treatments) jointly; AR(1) model.  The estimate of the mean CCR for \textbf{Version 9} is \textbf{0.998}.}}

\label{m004.F-upweek-and-all14nweeks-random.upgrade-indic.6mo-roll.ar1.lam0.cf330}

\end{figure*}

The straightforward approach to deal with the identifiability problem is to model all treatments jointly in a \emph{full model}, and include covariates (PCFs) that explicitly encode each user's waiting time to adopt the treatment.  Pooled data from other treatments reduces the collinearity in the design matrices because \emph{different users play the role of early adopters in each release}.  This may be seen by computing the correlation between the same covariate vectors in the full model (as we did in reduced model); the new value of only 0.216 means that there is little overlap between sets of early adopters from one version to the next.  This allows for robust estimates of the early-adopter effect, i.e., of the values of the $n$-weeks-past-release coefficients, which in turn produces robust estimates of the treatment indicator coefficients and a realistic CATT estimate.  These results are given in Figure \ref{m004.F-upweek-and-all14nweeks-random.upgrade-indic.6mo-roll.ar1.lam0.cf330}. The estimated mean CCR value for Version 9 is \textbf{0.998}, a negligible causal effect.   

\begin{figure*}[t!]

\centering

\begin{minipage}{0.45\textwidth}

\centering

\hspace*{-7pt}\includegraphics[scale=0.5,trim=0cm 0cm 0cm 0cm,clip=true]{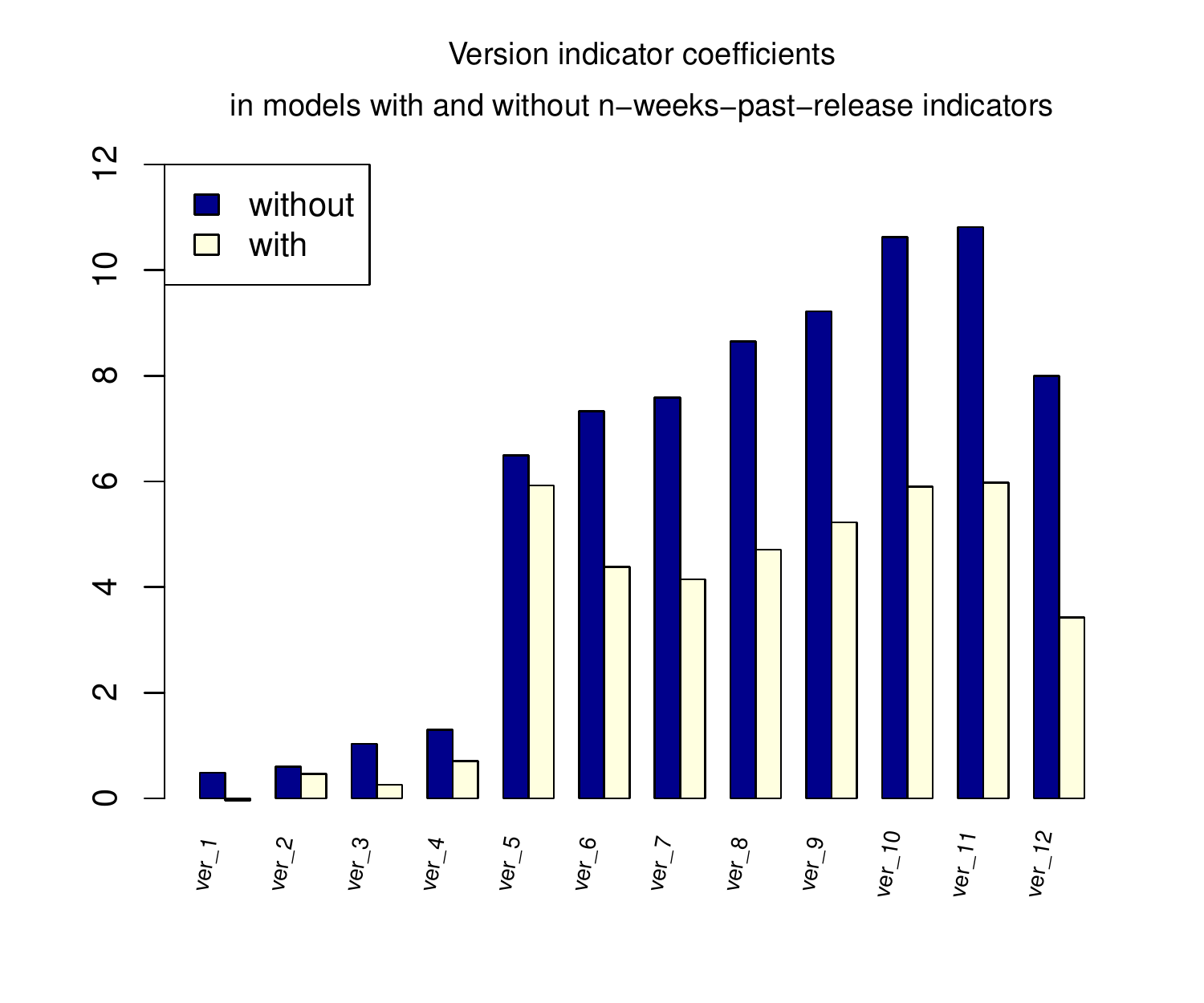}

\caption{\textit{Posterior means of the version coefficients in models with and without the $n$-weeks-past-release indicators. Including these indicators in the model eliminates the increasing trend in version coefficients and isolates the treatment effect.}}

\label{coef_ver_comparison}

\end{minipage}\hfill
\begin{minipage}{0.45\textwidth}

\centering

\hspace*{-7pt}\includegraphics[scale=0.5,trim=0cm 0cm 0cm 0cm,clip=true]{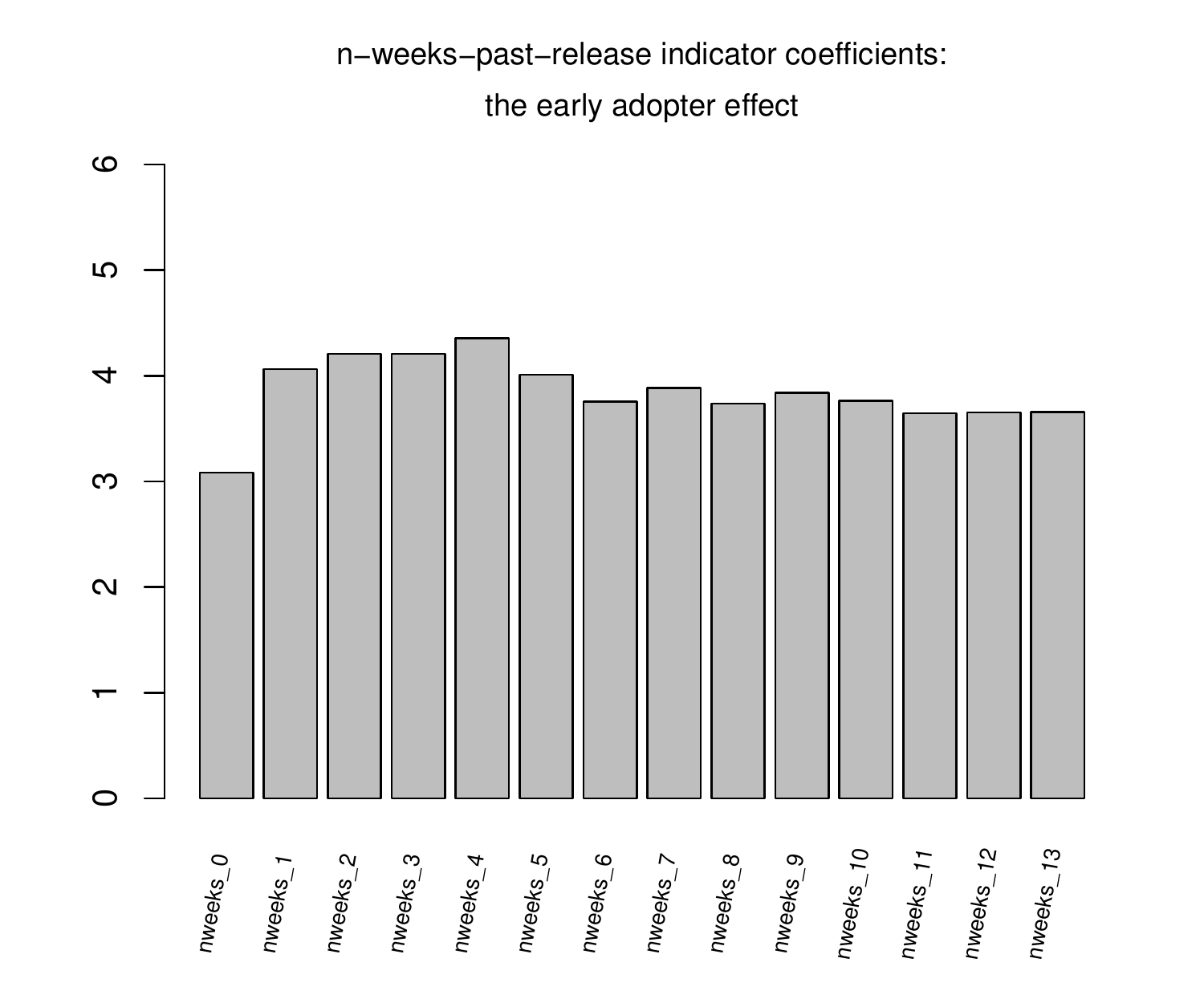}

\caption{\textit{Posterior means of the $n$-weeks-past-release indicators.  The magnitude of these coefficients shows that a significant portion of the response can be attributed to how long a user waited to adopt the treatment.}}

\label{coef_nweeks}

\end{minipage}

\end{figure*}

The early-adopter effect can also clearly be seen in Figures \ref{coef_ver_comparison} and \ref{coef_nweeks}.  Figure \ref{coef_ver_comparison} shows the posterior means of the components of $\boldsymbol{\mu}$ that correspond to the 12 versions (i.e., the version coefficients) resulting from two models:  first, a model that ignores the early-adopter effect and excludes the $n$-weeks-past-release indicators, and second, a model that does include these indicators. 
Note that in the first model, the version coefficients show an increasing trend, because the early-adopter effect is confounded with the treatment effect. (Version 12's coefficient does not follow the trend because Version 12 appears in only the last week in our study (week 52); in this case, the \textit{upgrade-week} indicator accounts for the first-week-adopter effect).  
The confounding occurs because, given a fixed 52-week time window, the fewer weeks a Product version has been on the market in the window, the larger its proportion of early adopters is as a fraction of its total users.  This has the effect of artificially driving up the version coefficient in the model, making it appear as if the version had a large positive effect on UA.  An obvious way to control for this effect is via the $n$-weeks-past-release indicators, whose posterior mean values are shown in in Figure \ref{coef_nweeks}.

\section{Model Selection and Validation} \label{model-selection}

Our goals in model selection and validation are twofold: (a) choose the single best model so that we can accurately estimate the CCR for other versions, and (b) make sure that our results are relatively robust to over-fitting.  Therefore, we study the sensitivity of our results to different AR orders, as well as investigate two additional model classes: the simpler flat model with a Gaussian error distribution given in equation (\ref{flat-model}), and a more complex model with a non-parametric error distribution.   Finally, we perform 5-fold cross-validation to verify that our results are not sensitive to some random subset of our users.

Our primary selection criterion for a best model is the \emph{the fit-of-the-treated}, namely the \emph{in-sample} root-mean-squared-error (RMSE) on the \emph{treatment group}.  The reasons for this are two-fold.  First, we are interested in computing CATT, namely the effect on the treatment group.  Second, we are (currently) not interested in predicting aggregate UA into the \emph{future}, nor are we interested in predicting how another totally different set of users would behave.  We are focused on accurately modeling \emph{our in-sample} 10.5M users (and the approximately 5M treated ones in particular) so that we can ascertain what \emph{their} behavior would have been like in the counterfactual world.  
The secondary selection criteria is scalability (computational feasibility).  Given that we are doing Bayesian inference over 10.5M users in an approximately 80-dimensional space, the ability to fit the model in a reasonable amount of time is important.  If a model takes an inordinate amount of time to fit, its performance advantage over competing models should be commensurate with this increased time and effort.

\subsection{AR Order Sensitivity} \label{ar-order-sensitivity}

Table \ref{rmseT_ar} gives the results for three identical hierarchical models in which we perturbed the AR order: $p = \{0,1,4\}$.  It is evident that the overall estimates of the causal effect --- as measured by the posterior mean of CCR for Version 9 --- are relatively insensitive to the AR order. However, the AR(4) model performs better than the AR(0) and AR(1) models (with minimal computational overhead), so we proceed with the AR(4) model in all subsequent analyses. \vspace*{-0.2in}

\begin{table}[t!]

\centering

\begin{tabular}{c|c|c|c|c|c}

& Error \\

Model Class & Distribution & AR order & RMSE Before & RMSE After & Mean CCR \\ 

\hline \hline

Hierarchical & Normal & 0 & 57.88 & 68.61 & 0.984 \\ 

\hline 

Hierarchical & Normal & 1 & 56.55 & 66.96 & 0.998 \\ 

\hline 

Hierarchical & Normal & 4 & 53.64 & 63.35 & 1.004 \\ 

\end{tabular}

\vspace*{0.1in}

\caption{\textit{RMSE $[\frac{1}{n_T}(\hat{\mathbf{Y}}_T - \mathbf{Y}_T)^2]^{1/2}$ for the \emph{treated} users over 7 weeks before and 7 weeks after the release of Version 9. The (posterior) mean CCR results are relatively insensitive to the AR order of the model, but more heterogeneity (i.e., a larger AR order) leads to a better fit of the treated.}}

\label{rmseT_ar}

\end{table}

\subsection{Sensitivity to Model Class} \label{sensitivity}

Having selected an AR(4) hierarchical model as our best candidate so far, we now study results from the simpler flat model (\ref{flat-model}) with a Gaussian error distribution and from a more complex model with a non-parametric error distribution.

\begin{figure*}[t!]

\centering

\includegraphics[scale=0.7,trim=0cm 0cm 0cm 0cm,clip=true]{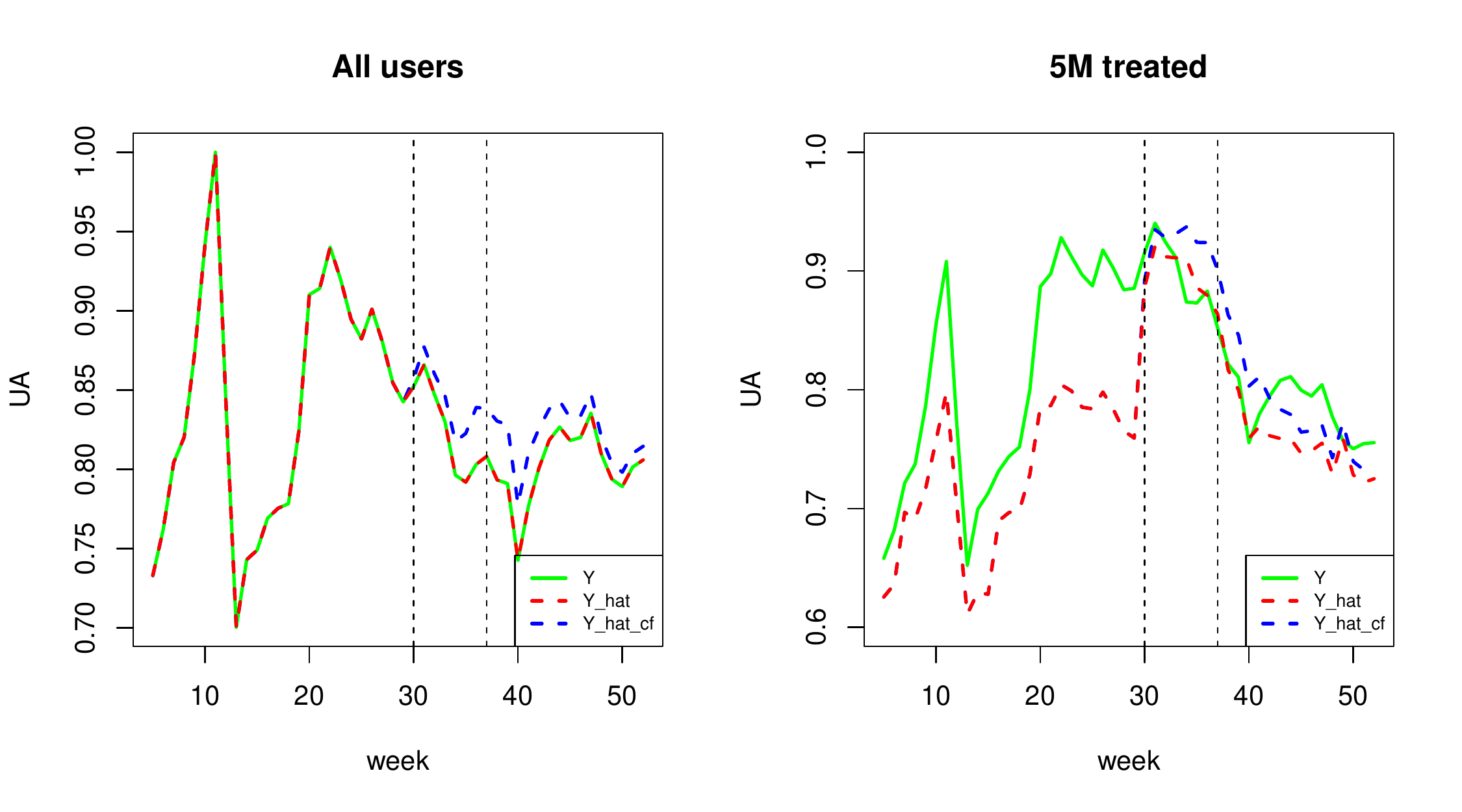}

\caption{\textit{Results for the \textbf{flat} AR(4) model; note the poor fit to the observed response of the treated in the right-hand plot. The estimate of the mean CCR for \textbf{Version 9} is \textbf{1.024}.}}

\label{m002.F-upweek-and-all14nweeks-random.upgrade-indic.6mo-roll.ar4.lam0.cf330}

\end{figure*}

\begin{itemize}

\item

First, we do not fit the flat OLS model because we consider it a sensible reflection of reality; it is not credible that all 10.5M users can be treated in a homogeneous fashion. However, it is a limiting case of our hierarchical model in which $\Sigma \rightarrow \textbf{0}$, and thus is a good candidate for analysis. In the last line of model (\ref{flat-model}) we assume an (improper) diffuse prior for all the parameters (proper diffuse priors yielded identical results).  

\item

Second, we fit the following Bayesian non-parametric (BNP) error model: for user $i = 1,\dots,n = $ 10,491,859,
\begin{align} \label{bnp-model}
\mathbf{y}_i &= \mathbf{f}_i\boldsymbol{\beta}_i + \mathbf{W}_i\boldsymbol{\gamma} + \boldsymbol{\varepsilon}_i \nonumber \\
( \boldsymbol{\varepsilon}_{i} \, | \, \theta_i, \nu ) &\sim \textbf{N}(\theta_i \, \mathbf{1}_{T-p}, \nu \, \textbf{I}_{T-p}) \nonumber \\
( \theta_i \, | \, \mathcal{P} ) &\sim \mathcal{P} \nonumber \\
\mathcal{P} &\sim \text{DP}(\alpha, \mathcal{P}_0) &\;\;\;\; \alpha=3 \text{ (fixed)} \nonumber \\
\mathcal{P}_0 &= \textbf{N}(0,\tau_0) &\;\;\;\; \tau_0=100 \text{ (fixed)} \nonumber \\
\boldsymbol{\beta}_i &\sim \textbf{N}(\boldsymbol{\mu},\mathbf{\Sigma}) \nonumber \\
(\nu, \boldsymbol{\mu},\mathbf{\Sigma}, \boldsymbol{\gamma}) &: \text{as in equation (\ref{primary-model})} \, .
\end{align}
Model (\ref{bnp-model}) is an expansion of our main hierarchical model (\ref{primary-model}), in that it allows the error distribution to have a (more realistic) non-parametric form, namely a Dirichlet process (DP) mixture of Gaussians. 
The idea is that the non-parametric DP (location) mixture form of the error distribution will do a better job fitting the small percentage of high UA users.  We fit model (\ref{bnp-model}) using the \emph{marginal Gibbs sampler} \citep{bda3}. (Varying $\alpha$ and $\tau_0$ across reasonable ranges had little effect on the results.)

The results for different model classes are shows in Table \ref{rmseT_modelclass}; we have included the Normal hierarchical model as well for comparison. Evidently, given the same covariates, all three models produce similar CCR estimates.  However, note that the flat model does an extremely poor job of fitting the treated users.  Figure \ref{m002.F-upweek-and-all14nweeks-random.upgrade-indic.6mo-roll.ar4.lam0.cf330} graphically illustrates this lack of in-sample fit to the treated.  Without the benefit of similar CCR estimates from the hierarchical model, its CCR estimate of 1.024 would be hard to believe given such a poor fit to the treated.  This illustrates the importance of taking user heterogeneity into account, especially when one is trying to estimate the causal effect on the treated.  

At the other end of the model complexity spectrum, Table \ref{rmseT_modelclass} shows that the BNP model fit slightly better but yielded a CCR estimate that is similar to that from the hierarchical Gaussian model: the posterior mean CCR moved from 1.004 to 1.011.  However, this came at a great computational cost: it took approximately 3 times the amount of clock time (with the same hardware) to fit the DP mixture model as it did the Gaussian hierarchical model with the same exact covariates.  Since the model with the non-parametric error distribution shows only minor differences in the CCR estimates, we argue that although its error distribution may be more realistic at the individual user level, it does not practically matter when considering the aggregate (and the mean) response.  As we mentioned previously, the error distribution for any single user is not Gaussian, but the lack of sensitivity to the exact form of error distribution at the aggregate level allows us to use the simpler hierarchical Gaussian model (\ref{primary-model}) with a reasonable degree of confidence.

\end{itemize}

\begin{table}[t!]

\centering

\begin{tabular}{l|l|c|c|c|c}

& \multicolumn{1}{c|}{Error} \\

Model Class & Distribution & AR order & RMSE Before & RMSE During & CCR \\ 

\hline \hline

Flat & Normal & 4 & 89.46 & 96.62 & 1.024 \\ 

\hline

Hierarchical & Normal & 4 & 53.64 & 63.35 & 1.004 \\ 

\hline  

Hierarchical & DP-mixture & 4 & 53.34 & 62.75 & 1.011 \\ 

\end{tabular}

\vspace*{0.1in}

\caption{\textit{RMSE $[\frac{1}{n_T}(\hat{\mathbf{Y}}_T - \mathbf{Y}_T)^2]^{1/2}$ for the \emph{treated} users over 7 weeks before and 7 weeks after the release of Version 9. Although the flat (homogeneous) model yields similar (posterior) mean CCR estimates, it exhibits a poor fit to the true response of the treated (see Figure \ref{m002.F-upweek-and-all14nweeks-random.upgrade-indic.6mo-roll.ar4.lam0.cf330}).  The non-parametric error model also produces similar results, but at a high computational cost.}}

\label{rmseT_modelclass}

\end{table}

\subsection{5-fold Cross Validation} \label{cross-validation}

Given that the hierarchical Gaussian AR(4) model is our chosen model, we performed 5-fold cross-validation to check its \textit{out-of-sample} (OOS) predictions. We fit 5 different variants of it using only 80\% of the users each time, and examined the fit and computed the CCR estimates on the remaining 20\% of the users. 

Table \ref{oosCATT} gives summaries of the posterior distribution of CCR for the treated users in each of the 5 subsets; each subset had approximately 2.1M out-of-sample (OOS) users (and approximately 1M OOS treated users). Each of the 95\% uncertainty bands includes 1, and the posterior mean CCR in each subset is similar to the value obtained using the whole set of users (see Table \ref{rmseT_modelclass}).

\begin{table}[t!]

\centering

\begin{tabular}{c|c|c|c|c}

& & \multicolumn{3}{c}{Posterior CCR} \\ \cline{3-5}

Subset & $n$ Treated Users & Lower 95\% & Mean & Upper 95\% \\ 

\hline

1 & 1,015,302 & 0.994 & 1.006 & 1.019 \\

2 & 1,014,554 & 0.997 & 1.009 & 1.023 \\

3 & 1,015,416 & 0.987 & 1.001 & 1.014 \\

4 & 1,013,886 & 0.994 & 1.007 & 1.020 \\

5 & 1,014,807 & 0.987 & 1.001 & 1.014

\end{tabular}

\caption{\textit{Simulated means and 95\% uncertainty bands for the CCR for Version 9, for five different held-out sets of treated users.}}

\label{oosCATT}

\end{table}

\begin{figure*}[t!]

\centering

\begin{minipage}{0.45\textwidth}

\begin{flushleft}

\hspace*{-7pt}\includegraphics[scale=0.5,trim=0cm 0cm 0cm 0cm,clip=true]{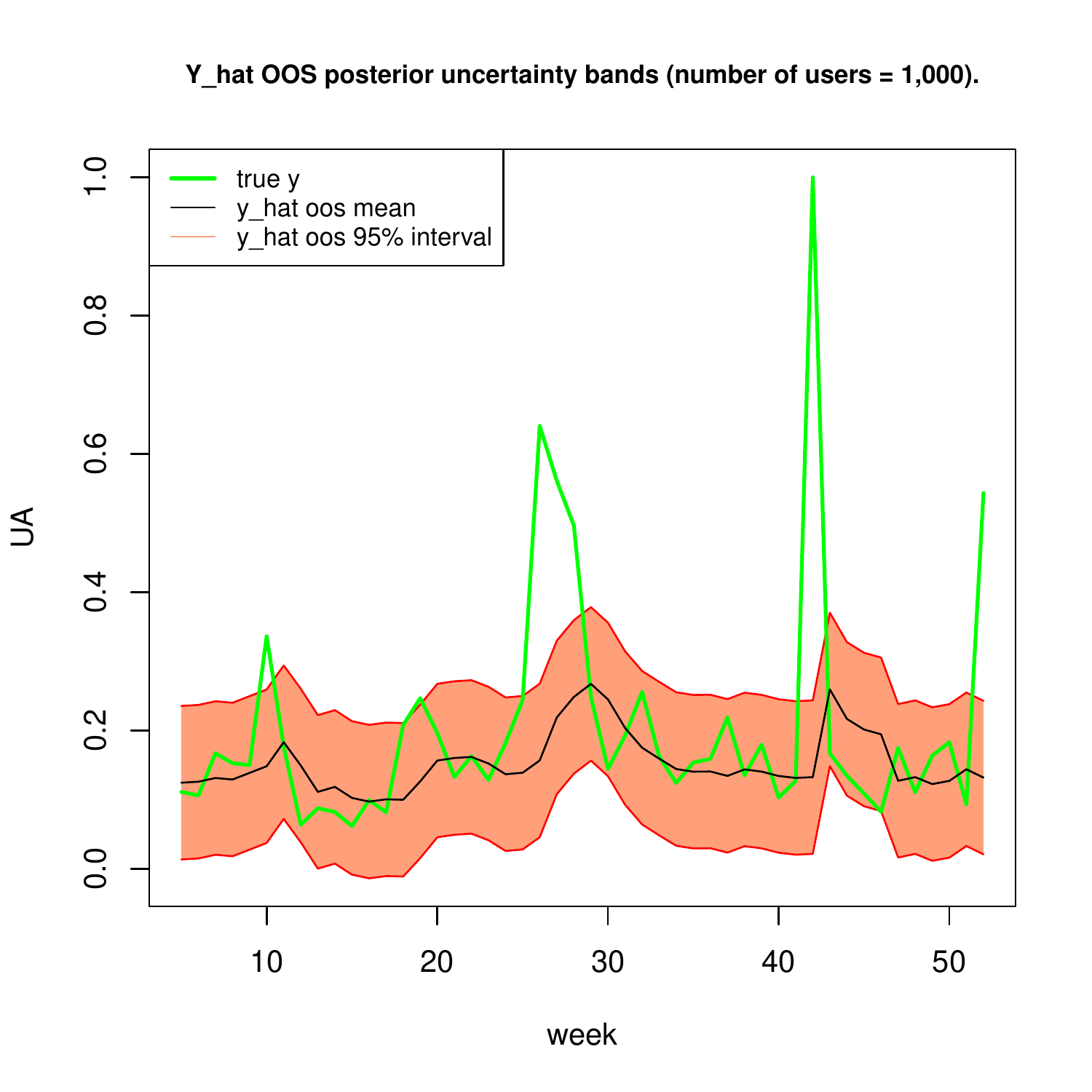}

\caption{\textit{Model fit results on 1,000 OOS users using the \underline{flat} model. This model, using the same covariates as its hierarchical counterpart in Figure \ref{m004.F-upweek-and-all14nweeks-random.upgrade-indic.6mo-roll.ar4.lam0.fold_1.1K}, cannot capture the nuances (e.g., non-normality) of the aggregate response for a small sample of users.}}

\label{m002.F-upweek-and-all14nweeks-random.upgrade-indic.6mo-
roll.ar4.lam0.fold_1.1K}

\end{flushleft}

\end{minipage}
\hspace{5mm}
\begin{minipage}{0.45\textwidth}

\begin{flushright}

\centering

\hspace*{-7pt}\includegraphics[scale=0.5,trim=0cm 0cm 0cm 0cm,clip=true]{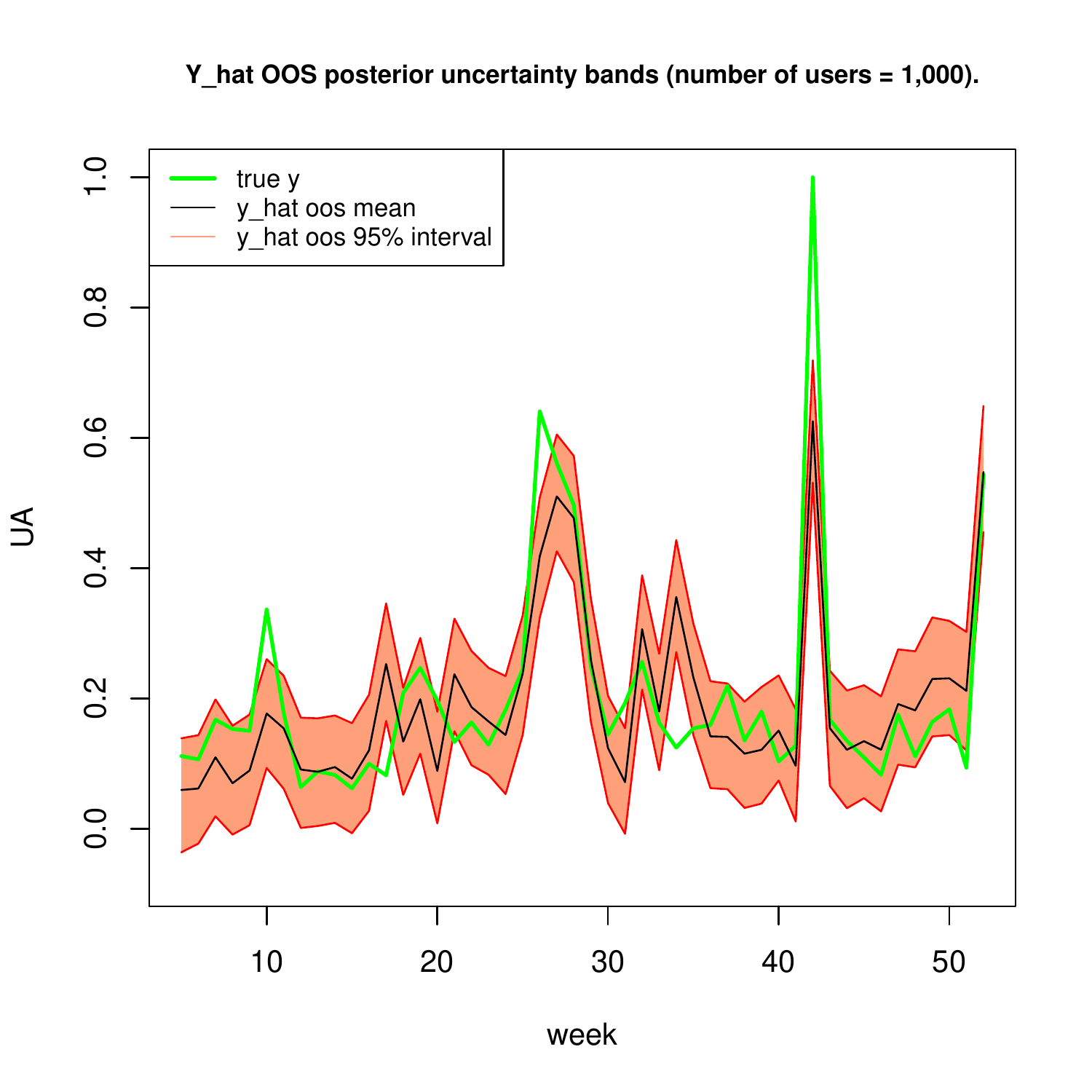}

\caption{\textit{Model fit results on 1,000 OOS users using our main \underline{hierarchical} model. Note how much better this model is able to capture the aggregate user response compared to the flat model in Figure \ref{m002.F-upweek-and-all14nweeks-random.upgrade-indic.6mo-
roll.ar4.lam0.fold_1.1K}.}}

\label{m004.F-upweek-and-all14nweeks-random.upgrade-indic.6mo-roll.ar4.lam0.fold_1.1K}

\end{flushright}

\end{minipage}

\end{figure*}

To further demonstrate the advantages of our main hierarchical model (\ref{primary-model}), we compare the ability of the flat and hierarchical models to predict the OOS aggregate response for a relatively small set of users, namely \emph{just 1,000 users} drawn randomly from the larger 2.1M OOS set. The two models under comparison have exactly the same covariates.  The results of this comparison are given in Figures \ref{m002.F-upweek-and-all14nweeks-random.upgrade-indic.6mo-
roll.ar4.lam0.fold_1.1K} and \ref{m004.F-upweek-and-all14nweeks-random.upgrade-indic.6mo-roll.ar4.lam0.fold_1.1K}. Figure \ref{m002.F-upweek-and-all14nweeks-random.upgrade-indic.6mo-
roll.ar4.lam0.fold_1.1K} illustrates the fit of the flat model on the set of 1,000 users, and Figure \ref{m004.F-upweek-and-all14nweeks-random.upgrade-indic.6mo-roll.ar4.lam0.fold_1.1K} performs the same task for the hierarchical model. The hierarchical model is vastly better at capturing the aggregate user response, and this improved fit is important in obtaining more accurate counterfactual estimates. As we noted in Section 4, one must \emph{first accurately fit the factual}, observed response \emph{before one can begin to predict the counterfactual} response, and doing so requires flexible (hierarchical) models that account for user heterogeneity.

\section{Summary of Results} \label{summary}

To highlight our main results, Table \ref{summary-table} shows the CCR estimates for two version releases, Version 9 and 10, using our chosen best model; we compare our results in this table to those obtained from a simple null model that contrasts the means of the treatment users 6 weeks before and 6 weeks after each release (results were similar with $\pm$ time windows other than 6). The null CCR estimates are wildly different from 1 on the low side, reflecting the magnitude of the early-adopter effect; the hierarchical estimates are close enough to 1 to allay any eBay fears of highly defective Product releases.

In conclusion, we have shown that careful consideration of long-term patterns of user behavior combined with flexible models yields causal estimates that are reasonable and stable. If the response to a treatment exhibits the characteristics of an early-adopter effect, then it is essentially impossible to isolate the causal effect of the treatment by analyzing a single treatment event in isolation. Attempting to do so (without including the appropriate PCFs) will yield treatment-effect estimates that are either unrealistic or unstable, because of model identifiablity issues. However, if the early-adopter effect exhibits a relatively consistent pattern from one treatment event to the next, we can estimate its contribution to the overall response by jointly modeling many similar treatment events simultaneously.  

Obtaining convincing CATT estimates requires doing a good job on the (in-sample) fit of the treated, which in turn requires flexible models that account for user heterogeneity. Hierarchical mixed-effects Bayesian models similar to the one presented here offer a good solution, for two reasons.  First, models in which each user has her/his own random effect naturally account for heterogeneity. Second, the hierarchical nature of the model makes it possible to borrow strength across multiple treatment events, permitting isolation of the 
early-adopter effect from the treatment effect.

\begin{table}[t!]

\centering

\begin{tabular}{c|c|c|c|c}

& & \multicolumn{3}{c}{Posterior CCR} \\ \cline{3-5}

Version & Model & Lower 95\% & Mean & Upper 95\% \\ 

\hline \hline

\ 9 & Null                              & 0.815         & 0.824    & 0.833         \\

\hline

\ 9 & \begin{tabular}{c} Hierarchical AR(4), \\ Normal Errors \end{tabular} & 0.998      & 1.004    & 1.010      \\

\hline

10              & Null                              & 0.709         & 0.720    & 0.731         \\

\hline

10              & \begin{tabular}{c} Hierarchical AR(4), \\ Normal Errors \end{tabular} & 1.022      & 1.028    & 1.035     

\end{tabular}

\caption{\textit{Summary of causal effect estimates for Versions 9 and 10.  The null model is a simple naive comparison of means for the treated group in a 6-week time window before and after the respective version release.}}

\label{summary-table}

\end{table}

\begin{appendices}

\section{CATT (Weak) Overlap Assumption}
\label{model-validation-overlap}
\begin{figure*}[ht]
\centering
\includegraphics[scale=0.4,trim=0cm 0cm 0cm 0cm,clip=true]{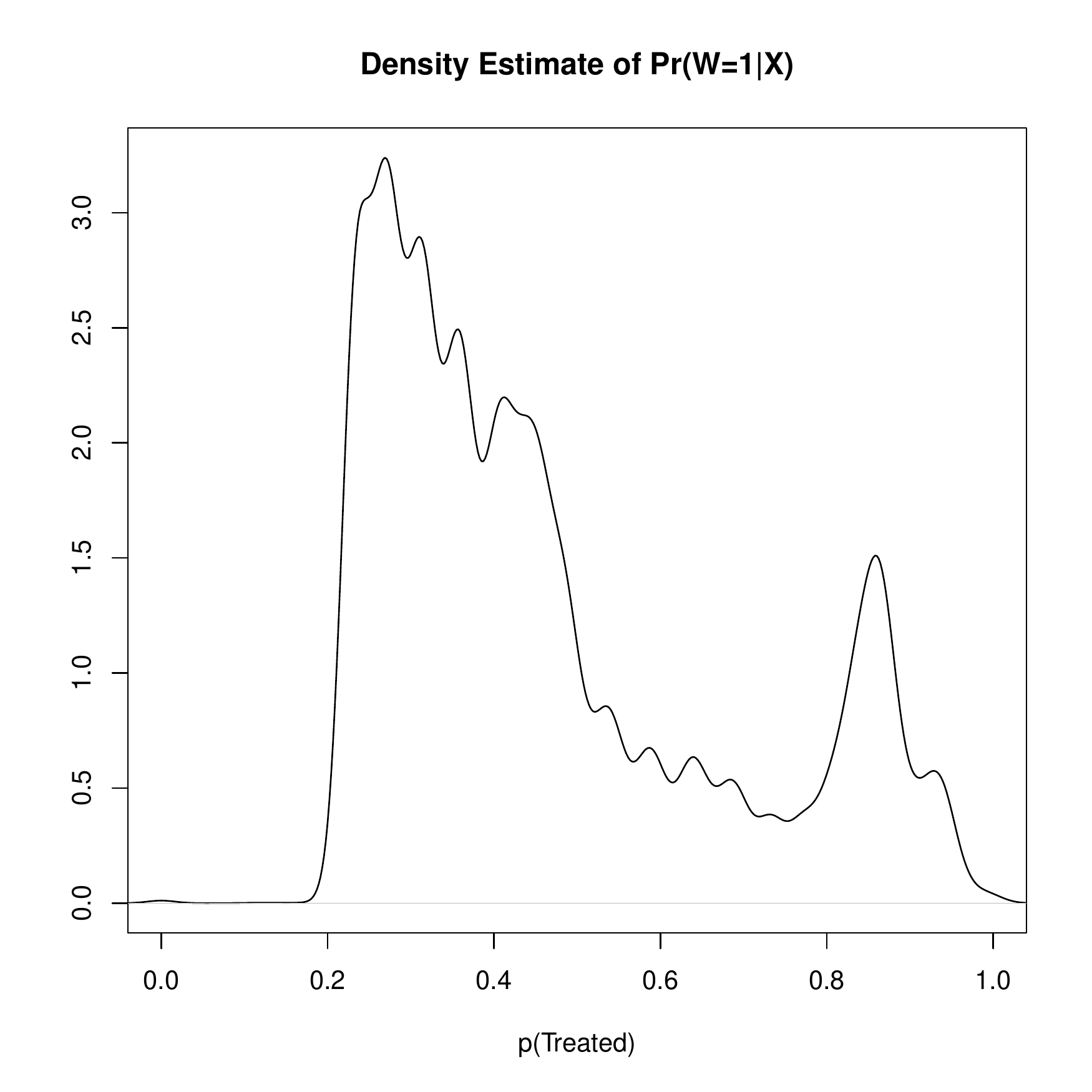}
\caption{\textit{Density estimate of the fitted propensity score $\hat{\pi}$ from a linear model, i.e. the probability of being in the treatment group conditional on the covariates.}}
\label{propensity-density}
\end{figure*}

For CATT estimates, it is sufficient that we verify $\pi \equiv \text{Pr}(Z=1|X) < 1$ \citep{heckman:ichi:todd:1997}. To do so we build a simple (linear) logistic regression model of the treated and control groups using our best covariates.  We collapse all of the $n$-weeks-past-release binary indicators into a single summed integer value in the range of $[0,13]$.  The estimated density plot of the resulting fitted probabilities $\hat{\pi}$ from the logistic regression model is shown in figure \ref{propensity-density}.  We can see that for the most part, $\hat{\pi} < 1$ is indeed true, except for a very few cases.  How few?  If we look at the quantiles of the fitted probability scores, we see the distribution in table \ref{propensity-quantiles}.  We see that the assumption holds for nearly all treated users; it fails in only 0.0069\% of cases.

\begin{table}[t]
\centering
\vspace{1cm}
\begin{tabular}{l|l|l|l|l|l|l|l|l}
20\%    & 30\%    & 50\%    & 70\%    & 90\%    & 95\%    & 96\%    & 99.90\% & 99.99\% \\
\hline
\hline
0.00004 & 0.00005 & 0.00012 & 0.15333 & 0.41694 & 0.60879 & 0.68938 & 0.95315 & 0.99991  \\
\end{tabular}
\vspace*{0.1in}
\caption{\textit{Quantiles of $\hat{\pi}$. Fewer than 0.01\% (0.0069\% to be exact) of the treated violate the $\hat{\pi} < 1$ condition.}}
\label{propensity-quantiles}
\end{table}

\pagebreak
\vspace{0.5cm}

\section{MCMC Sampling Equations}
\subsection{Hierarchical model: Gaussian Errors}
\label{mcmc-hier-eqn}
\singlespacing
\begin{align}
p(\boldsymbol{\beta}_i | \mathbf{y}_i, \mathbf{f}_i, \boldsymbol{\mu}, \mathbf{\Sigma}, \nu, \boldsymbol{\gamma}) 
    &= \textbf{N}(\boldsymbol{\beta}_i; \mathbf{m}_i, \mathbf{C}_i) \\
\mathbf{C}_i &= (\mathbf{\Sigma}^{-1} + \nu^{-1}\mathbf{f}_i^{'}\mathbf{f}_i)^{-1} \\
\mathbf{m}_i &= \mathbf{C}_i [\nu^{-1}\mathbf{f}_i^{'}(\mathbf{y}_i - \mathbf{W}_i \boldsymbol{\gamma}) + \mathbf{\Sigma}^{-1}\boldsymbol{\mu}]
\end{align}
\begin{align}
p(\boldsymbol{\mu}| \boldsymbol{\beta}_1,\boldsymbol{\beta}_2,\dots,\boldsymbol{\beta}_n, \mathbf{\Sigma}, \kappa_{\mu}) &= \textbf{N}(\boldsymbol{\mu}; \mathbf{a}, \mathbf{B}) \\
\mathbf{B} &= [(\kappa_{\mu}\mathbf{I})^{-1} + n \mathbf{\Sigma}^{-1}]^{-1} \\
\mathbf{a} &= \mathbf{B}(n \mathbf{\Sigma}^{-1} \bar{\boldsymbol{\beta}})
\end{align}
\begin{align}
p(\boldsymbol{\gamma} | \boldsymbol{\beta}_1,\boldsymbol{\beta}_2,\dots,\boldsymbol{\beta}_n, \nu, \mathbf{F}, \mathbf{Y}, \mathbf{W}, \kappa_{\gamma}) &= \textbf{N}(\boldsymbol{\gamma}; \mathbf{c}, \mathbf{D}) \\
\mathbf{D} &= [\nu^{-1}\mathbf{W}'\mathbf{W} + (\kappa_{\gamma}\mathbf{I})^{-1}]^{-1} \\
\mathbf{c} &= \nu^{-1}\mathbf{D}\mathbf{W}^{'}(\mathbf{Y}-\mathbf{F}\mathbf{B})\\
\textbf{W}'(\mathbf{Y}-\mathbf{F}\mathbf{B}) &= \sum_{i=1}^{n}\mathbf{W}_i^{'}(\mathbf{y}_i - \mathbf{f}_i \boldsymbol{\beta}_i) \;\;\;\; 
\text{and} \;\;\;\; \mathbf{W}'\mathbf{W} = \sum_{i=1}^{n}\mathbf{W}_i^{'}\mathbf{W}_i \nonumber
\end{align}
\begin{align}
p(\mathbf{\Sigma} | \boldsymbol{\mu}, \boldsymbol{\beta}_1,\boldsymbol{\beta}_2,\dots,\boldsymbol{\beta}_n) &=\text{Inv-Wishart}_{n+d+1}(\mathbf{S+I}) \\
\text{where } \mathbf{S} &= \sum_{i=1}^{n}(\boldsymbol{\beta}_i - \boldsymbol{\mu})(\boldsymbol{\beta}_i - \boldsymbol{\mu})' \nonumber
\end{align}
\begin{align}
p(\nu | \mathbf{Y}, \mathbf{F}, \boldsymbol{\beta}, \boldsymbol{\gamma}) &= 
	\text{IG}\Big[\nu; \frac{\epsilon + n(T-p)}{2}, \frac{\epsilon}{2} +
	\frac{1}{2}\sum_{i=1}^{n}(\mathbf{y}_i - \mathbf{f}_i\boldsymbol{\beta}_i - \mathbf{W}_i\boldsymbol{\gamma})^{'}(\mathbf{y}_i - \mathbf{f}_i\boldsymbol{\beta}_i - \mathbf{W}_i\boldsymbol{\gamma})\Big]
\end{align}

\doublespacing

\subsection{Hierarchical model: DP-Mixture Model for Errors}
The sampling algorithm is based on the \emph{marginal Gibbs sampler} \citep{bda3}, which separately updates the allocation of users to
clusters, and the cluster-specific parameters, as follows:
\begin{enumerate}
\item
\textbf{Update cluster allocation}:
\begin{itemize}
\item For $i=1,\dots,n$ users, compute the probability user $i$ belongs to one of the existing $k$ clusters, or to a totally new cluster:
\begin{itemize}
\item For $c=1,\dots,k$ clusters and the potentially new cluster $(k+1)$, compute the probability of each cluster as follows, where $n_c^{(-i)}$ is the number of users in cluster $c$ excluding user $i$, and $k^{(-i)}$ is the number of clusters that exist if user $i$ is not in any cluster:
\begin{numcases}{\mathbf{\pi}_i \equiv p(u_i = c) \propto}
n_c^{(-i)}\textbf{N}(\boldsymbol{\varepsilon}_i; \theta_c\mathbf{1}, \nu\mathbf{I}) &$c=1,\dots,k^{(-i)}$\label{existingLL}\\
\alpha \int \textbf{N}(\boldsymbol{\varepsilon}_i; \theta\mathbf{1}, \nu\mathbf{I}) \textbf{N}(\theta; 0, \tau_0) d\theta &$c=k^{(-i)}+1$\label{margLL}
\end{numcases}
\item Update the user's cluster membership by sampling from the $(k+1)$ dimensional multinomial: $c_i \leftarrow \text{Multinom}(\mathbf{\pi})$
\end{itemize}
\end{itemize}
\item
\textbf{Update cluster parameters}:
\begin{itemize}
\item For $c=1, \dots,k$ clusters, sample updated values of $\theta_c$: 
\begin{align}
\theta_c \sim p(\theta_c) \propto \textbf{N}(\theta_c;0,\tau)
\displaystyle\prod_{i : u_i \in c} \; \displaystyle\prod_{t=p+1}^{T}\textbf{N}(\varepsilon_{i,t}; \theta_c, \nu)
\label{ptheta}
\end{align}
\end{itemize}
\end{enumerate}
In detail, equation (\ref{existingLL}), the likelihood for an existing cluster $c$, is:
\begin{align}
n_c^{(-i)}\textbf{N}(\boldsymbol{\varepsilon}_i; \theta_c\mathbf{1}, \nu\mathbf{I}) &= n_c^{(-i)}\prod_{t=p+1}^{T}\textbf{N}( y_{i,t}-\mathbf{f}_{i,t}^{'}\boldsymbol{\beta}_i-\mathbf{W}_i\boldsymbol{\gamma}; \theta_c, \nu) \\
&= n_c^{(-i)}\prod_{t=p+1}^{T} \frac{1}{\sqrt{2\pi\nu}}\exp\Big\{-\frac{1}{2\nu}[\theta_c-(y_{i,t}-\mathbf{f}_{i,t}^{'}\boldsymbol{\beta}_i-\mathbf{W}_i\boldsymbol{\gamma})]^2\Big\}
\end{align} 
In detail, equation (\ref{margLL}), the marginal likelihood for a new cluster, is:
\begin{align}
\alpha &\int \textbf{N}(\boldsymbol{\varepsilon}_i; \theta\mathbf{1}, \nu\mathbf{I}) \textbf{N}(\theta; 0, \tau_0) d\theta 
=\alpha \int \prod_{t=p+1}^{T}\textbf{N}(\varepsilon_{i,t};\theta,\nu)\textbf{N}(\theta; 0, \tau_0) d\theta \\
& \hspace{4.5cm} = \frac{\alpha\sqrt{\nu}}{(\sqrt{2\pi\nu})^{(T-p)}\sqrt{(T-p)\tau_0+\nu}}
\exp\Big[-\frac{\sum_t \varepsilon_{i,t}^2}{2\nu}\Big]
\exp\Big\{\frac{\frac{\tau_0}{\nu}(T-p)^2(\bar{\varepsilon_i})^2}{2((T-p)\tau_0+\nu)}\Big\}
\end{align}
where $\bar{\varepsilon_i}$ is the mean residual value for user $i$ over $t=(p+1),\dots,T$, and $p$ is the AR order.

In detail, equation (\ref{ptheta}), is:
\begin{align}
\theta_c \sim p(\theta_c) &= \textbf{N}(\theta_c; a_c, b_c) \\
a_c &= b_c\Big(\frac{J\bar{\varepsilon_i}}{\nu}\Big) \;\;\;\; \text{where } J=(T-p)n_c \\
b_c &= \frac{1}{1/\tau_0 + J/\nu}
\end{align}

\end{appendices}

\bibliographystyle{chicago}

\bibliography{article}

\begin{thebibliography}{}

\bibitem[\protect\citeauthoryear{Angrist, Imbens, and Rubin}{Angrist
  et~al.}{1996}]{angrist1996}
Angrist, J.~D., G.~W. Imbens, and D.~B. Rubin (1996).
\newblock Identification of causal effects using instrumentral variables.
\newblock {\em Journal of the American Statistical Association\/}~{\em
  91\/}(1), 444--455.

\bibitem[\protect\citeauthoryear{Brodersen, Gallusser, Koehler, Remy, and
  Scott}{Brodersen et~al.}{2015}]{brodersen2015}
Brodersen, K.~H., F.~Gallusser, J.~Koehler, N.~Remy, and S.~L. Scott (2015).
\newblock Inferring causal impact using {B}ayesian structural time-series
  models.
\newblock {\em Ann. Appl. Stat.\/}~{\em 9\/}(1), 247--274.

\bibitem[\protect\citeauthoryear{Fisher}{Fisher}{1935}]{Fisher1935}
Fisher, R. (1935).
\newblock {\em Design of Experiments}.
\newblock Oxford and Boyd.

\bibitem[\protect\citeauthoryear{Gelman, Carlin, Stern, Dunson, Vehtari, and
  Rubin}{Gelman et~al.}{2014}]{bda3}
Gelman, A., B.~Carlin, H.~Stern, D.~Dunson, A.~Vehtari, and D.~Rubin (2014).
\newblock {\em Bayesian Data Analysis, Third Edition (Chapman \& Hall/CRC Texts
  in Statistical Science)}.
\newblock Chapman and Hall/CRC.

\bibitem[\protect\citeauthoryear{Heckman, Ichimura, and Todd}{Heckman
  et~al.}{1997}]{heckman:ichi:todd:1997}
Heckman, J.~J., H.~Ichimura, and P.~E. Todd (1997).
\newblock Matching as an econometric evaluation estimator: Evidence from
  evaluating a job training programme.
\newblock {\em The Review of Economic Studies\/}~{\em 64\/}(4), 605--654.

\bibitem[\protect\citeauthoryear{Hill}{Hill}{2011}]{Hill07bayesiannonparametric}
Hill, J.~L. (2011).
\newblock {{B}ayesian nonparametric modeling for causal inference}.
\newblock {\em Journal of Computational and Graphical Statistics\/}~{\em
  20\/}(1).

\bibitem[\protect\citeauthoryear{Holland}{Holland}{1986}]{holland86}
Holland, P.~W. (1986).
\newblock Statistics and causal inference.
\newblock {\em Journal of the American Statistical Association\/}~{\em 81},
  945--960.

\bibitem[\protect\citeauthoryear{Imbens}{Imbens}{2004}]{imbens2004nea}
Imbens, G. (2004).
\newblock {Nonparametric Estimation of Average Treatment Effects under
  Exogeneity: A Review}.
\newblock {\em Review of Economics and Statistics\/}~{\em 86\/}(1), 4--29.

\bibitem[\protect\citeauthoryear{Imbens and Rubin}{Imbens and
  Rubin}{2015}]{Imbens2015Book}
Imbens, G. and D.~Rubin (2015).
\newblock {\em Causal Inference for Statistics, Social, and Biomedical
  Sciences: An Introduction}.
\newblock Cambridge University Press.

\bibitem[\protect\citeauthoryear{Kang and Schafer}{Kang and
  Schafer}{2007}]{kang2007ddr}
Kang, J. and J.~Schafer (2007).
\newblock {Demystifying Double Robustness: A Comparison of Alternative
  Strategies for Estimating a Population Mean from Incomplete Data}.
\newblock {\em Statistical Science\/}~{\em 22\/}(4), 523.

\bibitem[\protect\citeauthoryear{Karabatsos and Walker}{Karabatsos and
  Walker}{2012}]{Karabatsos2012925}
Karabatsos, G. and S.~G. Walker (2012).
\newblock A {B}ayesian nonparametric causal model.
\newblock {\em Journal of Statistical Planning and Inference\/}~{\em 142\/}(4),
  925 -- 934.

\bibitem[\protect\citeauthoryear{McCandless, Gustafson, and Austin}{McCandless
  et~al.}{2009}]{McCandless2009}
McCandless, L.~C., P.~Gustafson, and P.~C. Austin (2009).
\newblock {B}ayesian propensity score analysis for observational data.
\newblock {\em Statistics in Medicine\/}~{\em 28\/}(1), 94--112.

\bibitem[\protect\citeauthoryear{Neyman}{Neyman}{1923}]{neyman93}
Neyman, J. (1923).
\newblock On the application of probability theory to agricultural experiments.
  {E}ssay on principles. section 9.
\newblock {\em Statistical Science\/}~{\em 5\/}(4), 465--472.
\newblock Translated and edited by Dabrowska, D.M. and Speed, T.P. (1993).

\bibitem[\protect\citeauthoryear{Reinsel}{Reinsel}{2013}]{reinsel13}
Reinsel, G.~C. (2013).
\newblock {\em Elements of Multivariate Time Series Analysis}.
\newblock New York: Springer.

\bibitem[\protect\citeauthoryear{Rosenbaum}{Rosenbaum}{1987}]{rosenbaum:1987e}
Rosenbaum, P. (1987).
\newblock Model-based direct adjustment.
\newblock {\em Journal of the American Statistical Association\/}~{\em 82},
  387--394.

\bibitem[\protect\citeauthoryear{Rosenbaum and Rubin}{Rosenbaum and
  Rubin}{1983}]{Rubin1983}
Rosenbaum, P.~R. and D.~B. Rubin (1983).
\newblock The central role of the propensity score in observational studies for
  causal effects.
\newblock {\em Biometrika\/}~{\em 70}, 41--55.

\bibitem[\protect\citeauthoryear{Rosenbaum and Rubin}{Rosenbaum and
  Rubin}{1984}]{rose:rubi:redu:1984}
Rosenbaum, P.~R. and D.~B. Rubin (1984).
\newblock Reducing bias in observational studies using subclassification on the
  propensity score.
\newblock {\em Journal of the American Statistical Association\/}~{\em 79},
  516--524.

\bibitem[\protect\citeauthoryear{Rosenbaum and Rubin}{Rosenbaum and
  Rubin}{1985}]{rosenbaum1985mj}
Rosenbaum, P.~R. and D.~B. Rubin (1985).
\newblock Constructing a control group using multivariate matched sampling
  methods that incorporate the propensity score.
\newblock {\em The American Statistician\/}~{\em 39}, 33--38.

\bibitem[\protect\citeauthoryear{Rubin}{Rubin}{1973}]{Rubin1973}
Rubin, D.~B. (1973).
\newblock Matching to remove bias in observational studies (corr: {V}30 p728).
\newblock {\em Biometrics\/}~{\em 29}, 159--183.

\bibitem[\protect\citeauthoryear{Rubin}{Rubin}{1974}]{rubin74}
Rubin, D.~B. (1974).
\newblock Estimating causal effects of treatments in randomized and
  nonrandomized studies.
\newblock {\em Journal of Educational Psychology\/}~{\em 66}, 688--701.

\bibitem[\protect\citeauthoryear{Rubin and Thomas}{Rubin and
  Thomas}{2000}]{rubi:thom:comb:2000}
Rubin, D.~B. and N.~Thomas (2000).
\newblock Combining propensity score matching with additional adjustments for
  prognostic covariates.
\newblock {\em Journal of the American Statistical Association\/}~{\em
  95\/}(450), 573--585.

\bibitem[\protect\citeauthoryear{Stuart}{Stuart}{2010}]{stuart2010}
Stuart, E.~A. (2010).
\newblock Matching methods for causal inference: A review and a look forward.
\newblock {\em Statistical Science\/}~{\em 25\/}(1), 1--21.

\end{thebibliography}

\end{document}